\documentclass[superscriptaddress,twocolumn,showpacs,preprintnumbers,amsmath,amssymb,prb]{revtex4-1}

\usepackage{graphicx}% Include figure files
\usepackage{dcolumn}% Align table columns on decimal point
\usepackage{bm}% bold math

\begin{document}

\preprint{Journal ref.: Phys.\ Rev.\ B \textbf{87}, 195124 (2013)}

\title{Electronic structure of hole-doped delafossite oxides CuCr$_{1-x}$Mg$_x$O$_2$}

\author{T.~Yokobori}
\author{M.~Okawa}
\author{K.~Konishi}
\author{R.~Takei}
\author{K.~Katayama}
\affiliation{Department of Applied Physics, Tokyo University of Science, Shinjuku, Tokyo 162-8601, Japan}

\author{S.~Oozono}
\author{T.~Shinmura}
\author{T.~Okuda}
\affiliation{Department of Electrical and Electronics Engineering, Kagoshima University, Kagoshima, Kagoshima 890-0065, Japan}

\author{H.~Wadati}
\affiliation{Department of Applied Physics and Quantum-Phase Electronics Center, University of Tokyo, Bunkyo, Tokyo 113-8656 Japan}

\author{E.~Sakai}
\author{K.~Ono}
\affiliation{Photon Factory, KEK, Tsukuba, Ibaraki 305-0801, Japan}

\author{H.~Kumigashira}
\affiliation{Photon Factory, KEK, Tsukuba, Ibaraki 305-0801, Japan}
\affiliation{PRESTO, Japan Science and Technology Agency, Chiyoda, Tokyo 102-0076, Japan}

\author{M.~Oshima}
\affiliation{Department of Applied Chemistry, University of Tokyo, Bunkyo, Tokyo 113-8656, Japan}

\author{T.~Sugiyama}
\author{E.~Ikenaga}
\affiliation{Japan Synchrotron Radiation Research Institute, Sayo, Hyogo 679-5198, Japan}
 
\author{N. Hamada}
\affiliation{Department of Physics, Tokyo University of Science, Noda, Chiba 278-8510, Japan}

\author{T. Saitoh}
\email[Author to whom correspondence should be addressed.\\ Electronic address: ]{t-saitoh@rs.kagu.tus.ac.jp}
\affiliation{Department of Applied Physics, Tokyo University of Science, Shinjuku, Tokyo 162-8601, Japan}

\date{\today}

\begin{abstract}
We report the detailed electronic structure of a hole-doped delafossite oxide CuCr$_{1-x}$Mg$_x$O$_2$
($0 \leq x \leq 0.03$) studied by photoemission spectroscopy (PES), soft x-ray absorption spectroscopy
(XAS), and band-structure calculations within the local-density approximation +$U$ (LDA+$U$) scheme.
Cr/Cu $3p$-$3d$ resonant PES reveals that the near-Fermi-level leading structure has primarily the Cr
$3d$ character with a minor contribution from the Cu $3d$ through Cu $3d$--O $2p$--Cr $3d$
hybridization, having good agreement with the band-structure calculations.
This indicates that a doped hole will have primarily the Cr $3d$ character.
Cr $2p$ PES and $L$-edge XAS spectra exhibit typical Cr$^{3+}$ features for all $x$, while the Cu
$L$-edge XAS spectra exhibited a systematic change with $x$.
This indicates now that the Cu valence is monovalent at $x=0$ and the doped hole should have
Cu $3d$ character.
Nevertheless, we surprisingly observed two types of charge-transfer satellites that should be attributed
to Cu$^+$ ($3d^{10}$) and Cu$^{2+}$ ($3d^{9}$) like initial states in Cu $2p$-$3d$ resonant PES
spectrum of at $x=0$, while Cu $2p$ PES spectra with no doubt shows the Cu$^+$ character even for
the lightly doped samples.
We propose that these contradictory results can be understood by introducing not only the Cu $4s$ state,
but also finite Cu $3d$, $4d$--Cr $3d$ charge transfer via O $2p$ states in the ground-state electronic
configuration.
\end{abstract}


\maketitle

%-------------------- INTRODUCTION --------------------
\section{\label{sec:Int}Introduction}

The search for new sustainable energy resources, including new innovations, is an urgent issue in
modern societies.
Thermoelectricity is one of the promising candidates because there exists so much waste heat
that could be recovered without sacrificing environmental costs.
Delafossite-type oxides Cu$M$O$_2$ ($M$ = trivalent cation) have considerable potential for
thermoelectric materials \cite{Okuda05} because of their layered structure of edge-shared
$M$O$_6$ octahedrons that is very similar to the one in thermoelectric NaCoO$_2$.\cite{Terasaki97}
Hole-doped CuCr$_{1-x}$Mg$_x$O$_2$ is a member of this family, being a candidate for
a future thermoelectrode.
In CuCrO$_2$, $3d^3$ electrons of the Cu$^{3+}$ ions under the pseudo-$O_h$ local symmetry
fill up the narrow Cr $3d$ $t_{2g\uparrow}$ band, which is the conterpart of the Co $3d$
$t_{2g}$ band filled by six electrons in NaCoO$_2$.
Hence, as in Na$_x$CoO$_2$, a rapid change in the density of states (DOS) at the Fermi level
($E_F$) (Ref.\ \onlinecite{Takeuchi04}) may be realized near the $t_{2g}$ band edge in the
hole-doped system CuCr$_{1-x}$Mg$_x$O$_2$,\cite{Note1}
because the Cr $3d$ band is expected to be at the top of the valence band in terms of a comparison
of the charge-transfer energy of the Cr$^{3+}$ ion and that of the Cu$^+$ ion.\cite{Iwasawa06}
More precisely, in $k$-resolved electronic structure, this situation would correspond to the pudding-mold
band structure that yields a large thermopower $S$ in Na$_x$CoO$_2$.\cite{Kuroki07}
As a consequence, a combination of a large $S$ and the highest electrical conductivity $\sigma$ among
delafossite oxides\cite{Nagarajan01} may be able to produce a large thermoelectric figure of merit
$Z=S^2 \sigma / \kappa$ ($\kappa$: thermal conductivity) in the present system.

Aside from thermoelectricity, Cu$M$O$_2$ has various interesting physical properties both in fundamental
and applicational terms.
A former example is multiferroic oxides CuFeO$_2$ (Ref.\ \onlinecite{Kimura06}) and the present compound
CuCrO$_2$ (Ref.\ \onlinecite{Seki08}) as well, whereas an important finding for the latter was a $p$-type
transparent conducting oxide (TCO); the $n$-type TCO's such as In$_2$O$_3$, SnO$_2$, or ZnO based
ones had been realized earlier,\cite{Hamberg86} yet the $p$-type counterpart was more difficult.
A delafossite CuAlO$_2$ was the first $p$-type TCO with high carrier mobility and a wide band gap.\cite{Kawazoe97} 
From the view point of the near-$E_F$ electronic structure, this was accomplished by hole doping into
a wide gap Cu$^+$ oxide, which has the $d^{10}$ closed shell.\cite{Kawazoe97}
Hence, the top of the valence band was expected to have the Cu $3d$ character with some O $2p$
one due to hybridization.

The electronic structure of CuCrO$_2$ has been investigated both theoretically and experimentally in the context
of TCO,\cite{Scanlon09,Arnold09, Hiraga11} or of thermoelectric/multiferroic materials.\cite{Maignan09}
Along the conventional strategy for TCO, the top of the valence band is expected to have mainly the Cu $3d$
character, whereas it would be desirable to have mainly the  Cr $3d$ character for better thermoelectric properties
as mentioned before.
On this point, reported first-principles band-structure calculations are still controversial;
Scanlon \textit{et al.} reported that the Cr $3d$ partial DOS has the maximum peak at the same
energy as the maximum peak of the Cu $3d$ partial DOS and negligibly small Cr $3d$ partial DOS at
the top of the valence band.\cite{Scanlon09}
In contrast, Maignan \textit{et al.} reported considerable Cr $3d$ partial DOS at the top of the valence
band,\cite{Maignan09} and a recent study by Hiraga \textit{et al.} showed the Cr $3d$ partial DOS in a much
deeper energy.\cite{Hiraga11}
Experimental electronic structure of CuCrO$_2$ has been investigated by photoemission spectroscopy (PES),
x-ray absorption spectroscopy (XAS), and x-ray emission spectroscopy. 
In these studies, Scanlon \textit{et al.} and Arnold \textit{et al.} interpreted the development of the upper
part of the valence band with $x$ in CuAl$_{1-x}$Cr$_x$O$_2$ as a reconstruction of the Cu $3d$ bands in
stead of a development of the Cr $3d$ states, and concluded that the Cr $3d$ DOS minimally contributed to
the top of the valence band.\cite{Scanlon09,Arnold09}
However, magnetic and transport studies reported a close coupling of the doped holes by Mg substitution
and the spin of the Cr ions that suggested the mixed-valences state Cr$^{3+}$/Cr$^{4+}$,\cite{Okuda05,Ono07}
which in turn implies Cr $3d$ character at the top of the valence band in the parent compound CuCrO$_2$.

From the above overview, the electronic structure of CuCr$_{1-x}$Mg$_x$O$_2$, particularly near $E_F$,
has not been established yet.
In this paper, we performed a comprehensive study on the electronic structure of lightly hole-doped
CuCr$_{1-x}$Mg$_x$O$_2$ ($x$ = 0--0.03) by photoemission spectroscopy with various photon energies,
soft x-ray absorption spectroscopy, and band structure calculations using the local density approximation
+$U$ (LDA+$U$) method.

%-------------------- EXPERIMENT --------------------
\section{\label{sec:Exp}Experiment and Calculation}

Polycrystalline samples of CuCr$_{1-x}$Mg$_x$O$_2$ ($x$=0, 0.02, 0.03) were prepared by
the standard solid-state reaction.\cite{Okuda05}
Vacuum ultraviolet (VUV)-PES measurements in the range of the Cr/Cu 3$p$-$3d$ resonance
($h\nu = 40$--90 eV) were performed at BL-28A of the Photon Factory, KEK, using a SCIENTA SES-2000
electron analyzer. Hard x-ray PES (HX-PES) spectra taken with $h\nu =7940$ eV were measured at
BL47XU of SPring-8 using a SCIENTA R4000 electron analyzer.
XAS spectra of the Cr and Cu $L$ edge regions and Cu $2p$-$3d$ resonant soft x-ray PES (SX-PES)
spectra were measured at BL-2C of the Photon Factory, KEK, using a SCIENTA SES-2000 electron analyzer.
In order to obtain clean surface, we fractured the samples \textit{in situ} right before the measurements.
The fracturings and the measurements were done in ultrahigh vacuum, namely, about $2.0 \times 10^{-8}$ Pa
(VUV-PES, SX-PES, and XAS), about $1.2 \time 10^{-7}$ Pa (fractureing for HX-PES), and about
$2.5 \times 10^{-8}$ Pa (measurement for HX-PES), all at 300 K.
The intensity of the resonant PES spectra was normalized using photon current of the exit mirror.
The energy resolution was 30 meV (VUV-PES), 140 meV (SX-PES), and 250 meV (HX-PES).
All the Fermi-level ($E_F$) positions in the experiments were calibrated with Au spectra.

We also performed band-structure calculations with the full potential linearized augmented
plane-wave (FLAPW) method\cite{Andersen75,*Takeda79} in the LDA+$U$
scheme.\cite{Hohenberg64,*Kohn65,*Vosko80,Anisimov91,*Solovyev96,*Anisimov97}
For the effective Coulomb repulsion $U_{\text{eff}}=U-J$, relatively small values (2.0 eV for Cu and Cr)
were adopted.
The rhombic lattice parameters ($a=2.9760$ \AA, $c=17.1104$ \AA) were taken from Ref.\ \onlinecite{Poienar09}.
The plane-wave cut off energy was 653 eV for the wave function. We took 1313 $k$ points in
the irreducible Brillouin zone for the rhombohedral Brillouin zone.\cite{Maignan09}
Although the system is known to be antiferromagnetic,\cite{Okuda05} the magnetic structure was
assumed to be ferromagnetic\cite{Maignan09} because the detailed magnetic structure is not
experimentally well-determined.\cite{Poienar09}

%-------------------- RESULTS --------------------
\section{Results}

\subsection{Experimental valence-band electronic structure compared with band structure calculations}

%_________________________________________________________________________________________________________________________________
%Fig. 1
\begin{figure}[b]
	\begin{center}
 	\includegraphics[width=82mm,keepaspectratio]{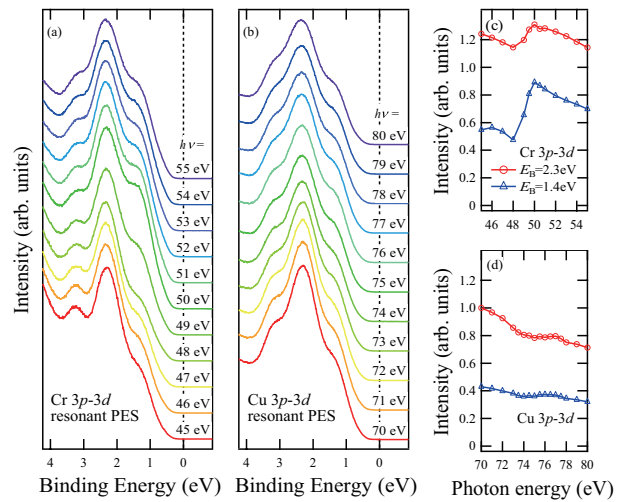}
  \caption{(Color online) Valence-band spectra of CuCr$_{0.98}$Mg$_{0.02}$O$_2$ taken with the photon energy around (a) the Cr 3$p$-$3d$ resonance region and (b) the Cu 3$p$-$3d$ resonance region. (c), (d) Constant initial-state spectra around the Cr 3$p$-$3d$ and Cu 3$p$-$3d$ resonances, respectively. Filled (open) symbols denote before (after) subtracting the background due to secondary electrons.}
\label{FIG_VUV-VB}
\end{center}
\end{figure}
%_________________________________________________________________________________________________________________________________

When the photon energy comes near the 3$p$-3$d$ (or 2$p$-3$d$) excitation threshold, resonant behaviors
appear in intensity of the valence-band photoemission due to the interference between the direct
($3d^n + h\nu \to 3d^{n-1} + e^-$) and indirect [$3d^n + h\nu \to 3p^53d^{n+1}$ (or $2p^53d^{n+1}$)
$\to 3d^{n-1} + e^-$] processes.
This is called 3$p$-3$d$ (or 2$p$-3$d$) resonant photoemission, which can be used to extract the 3$d$
contribution of a specific element to the valence band.
Figure~\ref{FIG_VUV-VB} shows the valence-band spectra of CuCr$_{0.98}$Mg$_{0.02}$O$_2$ taken with
a series of photon energy around the Cr 3$p$-$3d$ [Fig.\ \ref{FIG_VUV-VB}(a)] and the Cu 3$p$-$3d$
[Fig.\ \ref{FIG_VUV-VB}(b)] resonance.
One can easily observe that the intensity of the near-$E_F$ leading structures, namely, the shoulder at 1.4 eV
and the peak at 2.3 eV, systematically varies with incident photon energy. This intensity evolution is displayed
in Figs.\ \ref{FIG_VUV-VB}(c) and \ref{FIG_VUV-VB}(d) as the constant initial state (CIS) spectra at the binding
energy ($E_B$) of 1.4 eV [Fig.\ \ref{FIG_VUV-VB}(c)] and 2.3 eV [Fig.\ \ref{FIG_VUV-VB}(d)], respectively.
To remove the background intensity from the CIS spectra as taken (filled symbols),\cite{Li92} we also show
the CIS spectra after subtracting the background by the Shirley method (open symbols).\cite{Shirley72}

%_________________________________________________________________________________________________________________________________
%Fig. 2
\begin{figure}[t]
	\begin{center}
	\includegraphics[width=75mm,keepaspectratio]{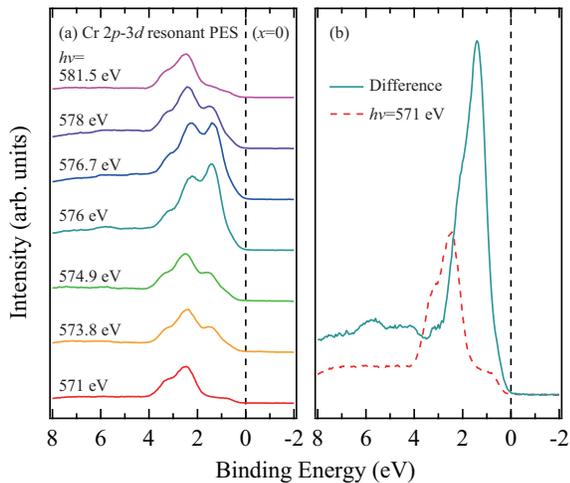}
	\caption{(Color online) (a) Valence-band spectra of CuCrO$_2$ taken with the photon energy around the Cr $2p$-$3d$ resonance region. The photon energies were determined by Cr $L$-edge XAS spectrum shown in Fig.\ \ref{FIG_Cr2p}. (b) On (576 eV) and off (571 eV) difference spectrum.}
\label{FIG_Cr2p3d}
\end{center}
\end{figure}
%_________________________________________________________________________________________________________________________________

%_________________________________________________________________________________________________________________________________
%Fig. 3
\begin{figure}[t]
	\begin{center}
 	\includegraphics[width=70mm,keepaspectratio]{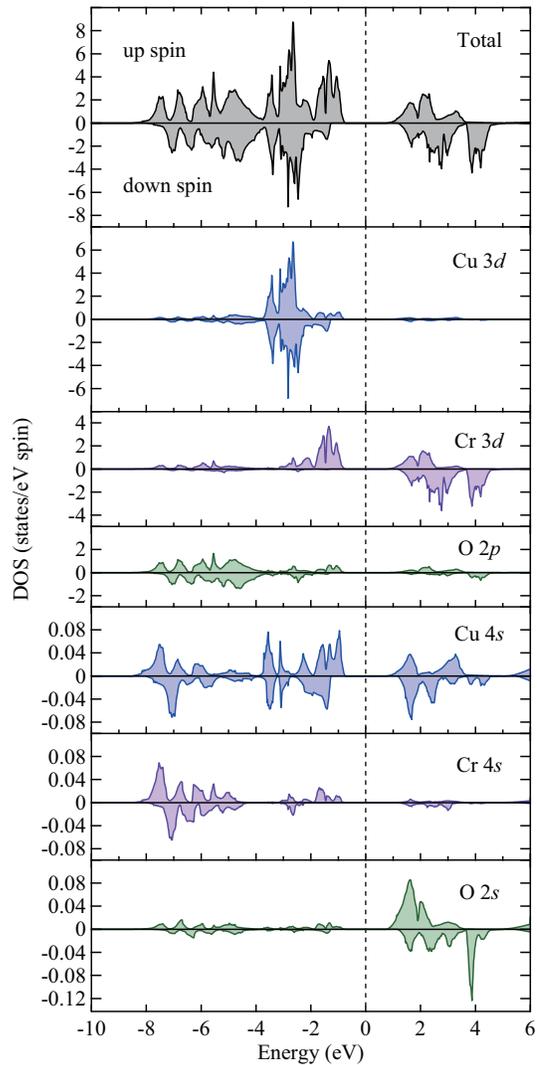}
 	\caption{(Color online) LDA+$U$ band structure calculations of CuCrO$_2$. $U_{\rm{eff}}$ was set to 2 eV both for Cr and Cu $d$ states.}
\label{FIG_LDAU}
\end{center}
\end{figure}
%_________________________________________________________________________________________________________________________________

Figure \ref{FIG_VUV-VB}(c) shows that the 1.4-eV shoulder exhibits a distinct resonance-type line shape\cite{Fano61}
with the maximum intensity at 50.0 eV, which is the Cr 3$p$-$3d$ resonance energy.\cite{Li92}
In contrast, the 2.3-eV peak shows a typical weak anti resonance-type line shape with a dip\cite{Fano61} at
the Cu 3$p$-$3d$ resonance energy 74.0 eV, as shown in Fig.\ \ref{FIG_VUV-VB}(d).\cite{Thuler82}
However, one also notices that a weak resonance of the 2.3-eV peak does exist at 50.0 eV and a tiny antiresonance
of the 1.4 eV at 74.0 eV.
These observations are clearly demonstrating that (1) the 1.4-eV shoulder includes a major contribution of the Cr $3d$
states with a minor contribution of the Cu $3d$ states and vice versa for the 2.3-eV peak, and (2) nevertheless there
exists sizable hybridization between the Cr $3d$ and Cu $3d$ states via O $2p$ states.
The major contribution of the Cr $3d$ states in the 1.4-eV shoulder is also confirmed by a $2p$-$3d$ resonant
PES measurement as shown in Fig.~\ref{FIG_Cr2p3d}.
Figure \ref{FIG_VUV-VB}(a) demonstrates that the 1.4-eV shoulder at $h\nu =571$ eV (off resonance) rapidly
grows to an intense peak at $h\nu =576$ eV (on resonance) with increasing photon energy.
Accordingly, the on-off difference spectrum, representing the Cr $3d$ partial DOS, has a sharp peak at 1.4 eV (Panel (b)).

From the above results, the schematic energy diagram is that the Cr $3d$ is at the top of the valence band,
the next is Cu $3d$, and then O $2p$ states come in the order of binding energy. 
This conclusion is different from recent PES\cite{Arnold09} or optical\cite{Hiraga11} studies, both of which
concluded that the Cu $3d$ states are located at the top of the valence band.
The origin of this difference will be discussed later in relation to band structure calculations.
The present result is reasonable also from the viewpoint of the O $2p-$TM $3d$ charge transfer energy $\Delta$
because the location of the Cr $3d$ states and the Cu $3d$ states, hybridizing with each other via O $2p$ states
in this compound, is governed by the difference of $\Delta_{\text{Cr}^{3+}}$ and
$\Delta_{\text{Cr}^{+}}$,\cite{Iwasawa06} and $\Delta_{\text{Cr}^{3+}}$ would be larger than
$\Delta_{\text{Cr}^{+}}$ even considering the different valence and local configurations.\cite{Fujimori93, Saitoh95}

In order to analyze the valence-band electronic structure in more detail, we performed LDA+$U$ band-structure
calculations.
Figure~\ref{FIG_LDAU} shows the result of our LDA+$U$ calculations.
The Cu $3d$ partial DOS has intense peaks between $-2.5$ and $-4.0$ eV with small Cr $3d$ partial DOS in this
range, whereas the Cr $3d$ partial DOS exhibits a considerably large peak centered at about $-1.5$ eV, distributed
from the top of the valence band to $-2.5$ eV with small Cu $3d$ partial DOS in this range.
Here, it is noted that the calculated Cr $3d$ partial DOS has good agreement with the experimental Cr $3d$
spectral weight in Fig.\ \ref{FIG_Cr2p3d}(b).
The O $2p$ bands are mainly located below the Cr and Cu $3d$ bands, from $-4$ to $-8$ eV. All the $s$ states,
Cu $4s$, Cr $4s$ and O 2$s$, show very small DOS in the entire energy range.
The present calculation, particularly on the location of the Cu/Cr $3d$ partial DOS, agrees well with the experimental
result shown in Fig.~\ref{FIG_VUV-VB} and the interpretation/prediction using the difference of $\Delta_{\text{Cr}^{3+}}$
and $\Delta_{\text{Cr}^{+}}$ as well.\cite{Iwasawa06}

%_________________________________________________________________________________________________________________________________
%Fig. 4
\begin{figure}[t]
	\begin{center}
 	\includegraphics[width=70mm,keepaspectratio]{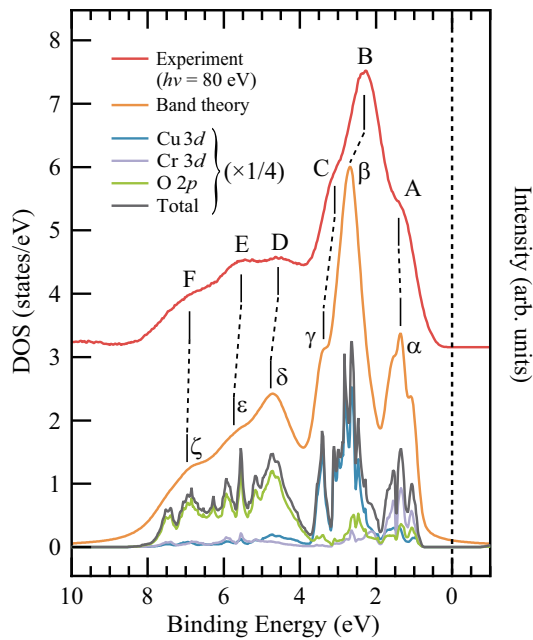}
 	\caption{(Color) Comparison between the calculated DOS of CuCrO$_2$ and the valence-band spectrum of  CuCr$_{0.98}$Mg$_{0.02}$O$_2$ taken at $h\nu=80.0$ eV.}
\label{FIG_Band}
\end{center}
\end{figure}
%_________________________________________________________________________________________________________________________________

The agreement between our experiment and calculation is demonstrated more clearly in Fig.~\ref{FIG_Band},
which shows a comparison between the experimental spectrum of CuCr$_{0.98}$Mg$_{0.02}$O$_2$ taken at
$h\nu=80$ eV and the calculated DOS.\cite{Note2}
A theoretical simulation curve has been constructed by broadening the cross-section-weighed total DOS with an
energy dependent Lorentzian function due to the lifetime effect and a Gaussian due to the experimental
resolution.\cite{Saitoh97,Iwasawa09,Yeh85,Note3}
This theoretical specctrum shows that the leading structure at the top of the valence band (labeled as $\alpha$) is
dominated by the Cr $3d$ states with a minor contribution of the Cu $3d$ states whereas the most intense peak
(labeled as $\beta$) primarily originates from the Cu $3d$ states. In both structures, appreciable O $2p$ DOS exist
as well because of large photoionization cross section.\cite{Yeh85}
One can see that the theoretical spectrum satisfactorily reproduces the experimental one and thus the experimental
structures A to F can be assigned to the theoretical structures $\alpha$ to $\zeta$, respectively. 

Our calculation agrees well with the calculation by Maignan \textit{et al.}\cite{Maignan09} while it is
different from Scanlon \textit{et al.}\cite{Scanlon09} or Hiraga \textit{et al.}\cite{Hiraga11}
However, we note that the spectrum by Scanlon \textit{et al.} and Arnold \textit{et al.} can simply be
interpreted by our calculation as a development of the Cr $3d$ states by Cr substitution for Al.\cite{Scanlon09,Arnold09}
Hence, we consider that their experiment is actually consistent with ours.
On the other hand, Hiraga \textit{et al.} consistently interpreted their optical absoption spectra using
their band structure calculations.\cite{Hiraga11} However, optical absorption spectroscopy is indirect
to probe the valence-band electronic structure because it gives the joint DOS.
While we (and Maignan \textit{et al.}) have assumed the ferromagnetic state in the calculations,
we believe that the different Cr $3d$ partial DOS does not come from the different magnetic structures
because both Scanlon \textit{et al.} and Hiraga \textit{et al.} have calculated antiferromagnetic states
by the same generalized gradient approximation +$U$ (GGA+$U$) method, resulting in the quite different
Cr $3d$ partial DOS's.
The differences in the two calculations probably originate from the fact that Scanlon \textit{et al.} adopted
theoretically optimized lattice parameters and Hiraga \textit{et al.} set the $U$ value for the Cu $3d$ states
to be zero.
Our result is also supported by another band structure calculation of CuAl$_{0.95}$Cr$_{0.05}$O$_2$ that
reported the same energetic order of the Cr and the Cu $3d$ states as ours, namely, the Cr $3d$ states come
to the top of the valence band by Cr doping.\cite{Kizaki05}

\subsection{Cu and Cr valence}

%_________________________________________________________________________________________________________________________________
%Fig. 5
\begin{figure}[t]
	\begin{center}
	\includegraphics[width=82mm,keepaspectratio]{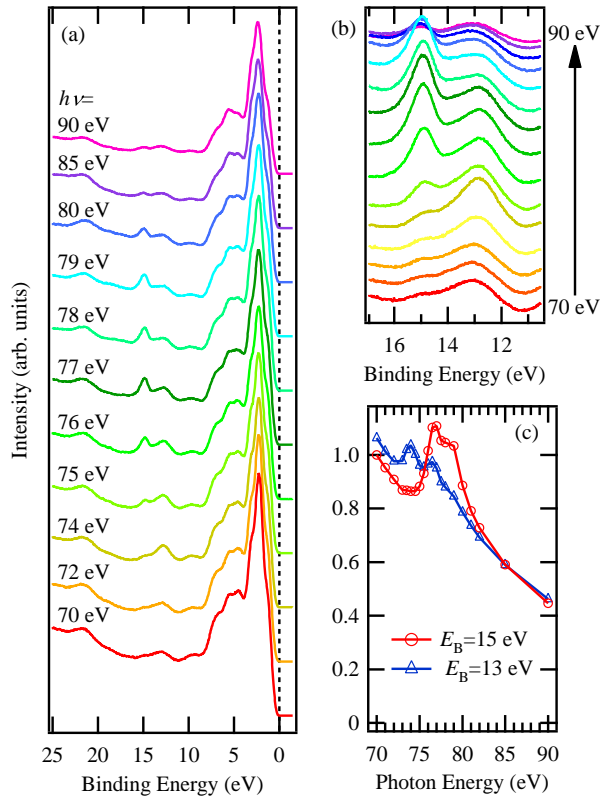}
	  \caption{(Color oneline) (a) Wide-range valence-band spectra of CuCr$_{0.98}$Mg$_{0.02}$O$_2$ in the energy range of the Cu satellite resonances. (b) Detailed satellite structures in Panel (a). (c) CIS spectrum of the two satellite peaks at the binding energy ($E_B$) of 13 and 15 eV.}
\label{FIG_VUV-Sat}
\end{center}
\end{figure}
%_________________________________________________________________________________________________________________________________

Figure \ref{FIG_VUV-Sat}(a) shows the valence-band photoemission spectra of CuCr$_{0.98}$Mg$_{0.02}$O$_2$
taken across the resonant energies of the Cu satellite structures.
There can be observed two distinct satellite peaks at the binding energy of 13 and 15 eV, which have their
maximums at the photon energy of 74 and 77 eV, respectively, as shown in Fig.\ \ref{FIG_VUV-Sat}(b).
These numbers are in very good agreement with the reported satellite peaks in CuO (12.5--12.9 eV) and
Cu$_2$O (15.3 eV), which have mainly $3d^8$ and $3d^84s$ final-state character,
respectively.\cite{Thuler82,Ghijsen90,Shen90}
The 15-eV satellite peak has also been observed in Al $K\alpha$ XPS spectra of CuAlO$_2$ and CuCrO$_2$.\cite{Arnold09}
Figure \ref{FIG_VUV-Sat}(c) shows the CIS spectra of these two satellite peaks.
The CIS profiles of the satellites again well reproduce those of CuO and Cu$_2$O, respectively, including
the two-peak structure due to 3$p_{3/2}$ and 3$p_{1/2}$ splitting.\cite{Thuler82,Ghijsen90}
All these results indicate that the doped hole in CuCr$_{0.98}$Mg$_{0.02}$O$_2$ produces Cu$^{2+}$ ions,
namely, holes will be doped into the Cu sites.
However, this observation seems to be incompatible with the result that the top of the valence band has mainly
the Cr $3d$ character, demonstrated in Figs.\ \ref{FIG_VUV-VB}--\ref{FIG_Band}.
Moreover, the 13-eV satellite due to Cu$^{2+}$ seems to be too intense for only 3\% doping of Mg, which
corresponds to 3\% Cu$^{2+}$ ions.

%_________________________________________________________________________________________________________________________________
%Fig. 6
\begin{figure}[t]
   \begin{center}
   \includegraphics[width=80mm, keepaspectratio]{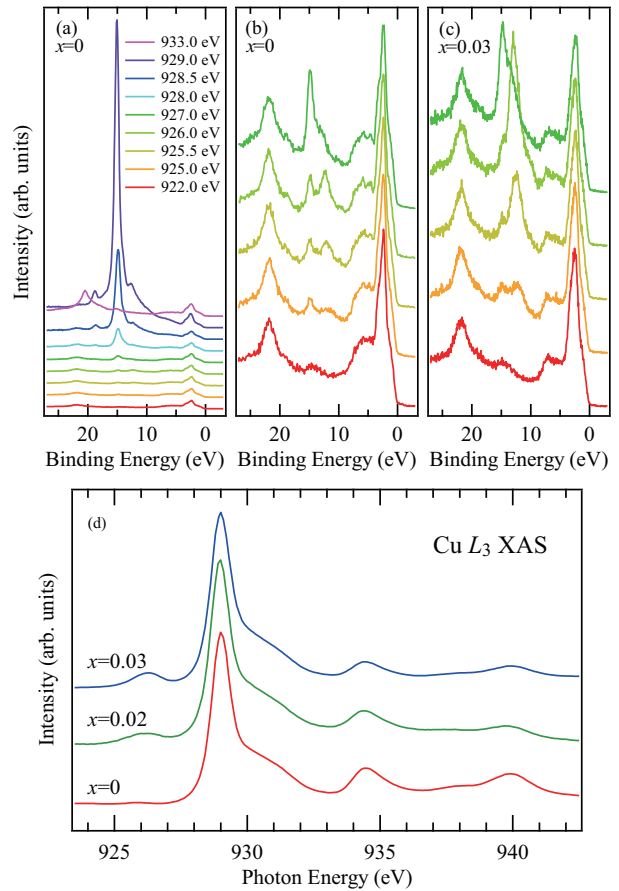}
   \caption{(Color) Valence-band spectra of CuCr$_{1-x}$Mg$_x$O$_2$ ($x=$ 0, 0.03) taken with the photon energy around the Cu $2p$-$3d$ resonance region. (a) Cu $2p$-$3d$ resonant spectra of the $x=0$ sample. (b) Same as (a) around the pre-peak energy region before the giant resonance. (c) Cu $2p$-$3d$ resonant spectra of $x=0.03$ sample in the same energy region as (b). (d) Cu $L_3$ XAS spectra of CuCr$_{1-x}$Mg$_x$O$_2$ ($x=$ 0, 0.02, 0.03).}
   \label{FIG_Cu2p3d}
   \end{center}
   \end{figure}
%_________________________________________________________________________________________________________________________________

To confirm this observation, we performed Cu $2p$-$3d$ resonant photoemission spectroscopy measurements,
as shown in Fig.\ \ref{FIG_Cu2p3d}. The excitation energies were determined by Cu $L_3$ XAS spectra shown
in Fig.\ \ref{FIG_Cu2p3d}(d).\cite{Note4}
Figure \ref{FIG_Cu2p3d}(a) shows the valence-band spectra of the $x=0$ sample taken in the Cu $2p$-$3d$
resonance region.
The giant resonance peak at 15 eV is due to Cu$^{+}$ ions as seen in Fig.\ \ref{FIG_VUV-Sat} and as reported
for Cu$_2$O.\cite{Tjeng92}
Figures \ref{FIG_Cu2p3d}(b) and \ref{FIG_Cu2p3d}(c) show the spectra taken at the photon energies before the
giant resonance develops.
In Fig.\ \ref{FIG_Cu2p3d}(c), the $x=0.03$ spectrum shows the distinct 13-eV resonant peak of Fig.\ \ref{FIG_VUV-Sat} at
$h\nu=926.0$ eV that corresponds to the photon energy of the pre-peak structure in Fig.\ \ref{FIG_Cu2p3d}(d).
This hump has been observed in some Cu$_2$O (Refs.\ \onlinecite{Hulbert84,Grioni89-1}), CuAlO$_2$
(Ref.\ \onlinecite{Aston05}) and CuCrO$_2$ (Ref.\ \onlinecite{Arnold09}) but has not been observed in pure
Cu$_2$O,\cite{Grioni92} and it is accordingly interpreted as $\underline{2p}3d^{10}$ final state due to
Cu$^{2+}$ impurity,\cite{Grioni92,Aston05,Arnold09} where $\underline{2p}$ denotes a core hole of the Cu $2p$ level.
Therefore, both the Cu $2p$-$3d$ resonant photoemission and the Cu $L_3$ XAS spectra of the $x=0.03$ sample
clearly demonstrate the Cu $3d$ nature of the doped holes observed in Fig.\ \ref{FIG_VUV-Sat}.

Surprisingly, however, Fig.\ \ref{FIG_Cu2p3d}(b) shows that the $x$=0 sample, too, has the 13-eV satellite.
This can never be due to Cu$^{2+}$ impurity because the Cu $L_3$ XAS spectrum has no appreciable
prepeak [see Fig.\ \ref{FIG_Cu2p3d}(d)].
Here, we noted that the very slight modulation from the baseline at the prepeak of the $x$=0 spectrum cannot
explain the large 13-eV resonance peak because the Cu$^{2+}$ impurity concentration in the $x=0.03$ sample,
if exists, can be estimated to be a few percent at most by a comparison with the reported relation between the
concentration and the prepeak intensity in CuAlO$_2$.\cite{Aston05}
Therefore, it can be undoubtedly concluded that some kind of $3d^9$ state that does not originate from Cu$^{2+}$
impurities, should exist even in the pure CuCrO$_2$, and based on this fact, one may further go beyond the $x=0$
case, and arrive at the idea that the whole portion of a doped hole may not necessarily go into a Cu site even because
the 13-eV satellite is observed.

%_________________________________________________________________________________________________________________________________
%Fig. 7
\begin{figure}[t]
	\begin{center}
	\includegraphics[width=80mm,keepaspectratio]{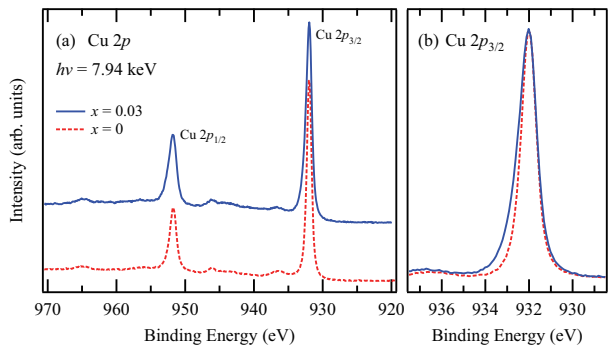}
	\caption{(Color online) (a) Cu $2p$ core-level HX-PES spectra of CuCr$_{1-x}$Mg$_x$O$_2$ ($x=$ 0, 0.03). (b) Comparison of the $2p_{3/2}$ peak of $x=0$ and 0.03. Note that the background due to secondary electrons was subtracted by the Shirley method.}
\label{FIG_Cu2p}
\end{center}
\end{figure}
%_________________________________________________________________________________________________________________________________

Figure \ref{FIG_Cu2p} shows Cu $2p$ core-level spectra of CuCrO$_2$ and CuCr$_{0.97}$Mg$_{0.03}$O$_2$.
The $x=0$ spectrum in Panel (a) is almost identical to the reported spectra of CuCrO$_2$ (Refs.\ \onlinecite{Arnold09,Le11})
and also CuAlO$_2$.\cite{Aston05}
There is no trace of structures at 934 eV due to the Cu$^{2+}$ state that, if exist, can easily be identified
as is the case of oxidized CuAl$_{1-x}$Zn$_x$O$_2$ or CuRh$_{1-x}$Mg$_x$O$_2$.\cite{Aston05,Le11}
A reported energy shift of the Cu $2p_{3/2}$ peak due to Mg doping\cite{Arnold09} was not observed and
the $x=0.03$ spectrum is almost identical to that of $x=0$, which is very similar to what was observed in
CuAl$_{1-x}$Zn$_x$O$_2$.\cite{Aston05}	
This fact raises doubt about the Cu $3d$ nature of a doped hole. Nevertheless, a small but important change
due to Mg doping can be observed in Fig.\ \ref{FIG_Cu2p}(b); the Cu $2p_{3/2}$ line shape becomes
asymmetrically broad. 
A Doniach-\v{S}unji\'c lineshape analysis\cite{Doniach70} has confirmed a large increase in asymmetry with
hole doping, which is reflecting an increase in metallicity of the system, particularly on the Cu sites.\cite{Note5}
Hence, this small change suggests the Cu $3d$ nature of a doped hole again.

%_________________________________________________________________________________________________________________________________
%Fig. 8
\begin{figure}[t]
	\begin{center}
	\includegraphics[width=67mm,keepaspectratio]{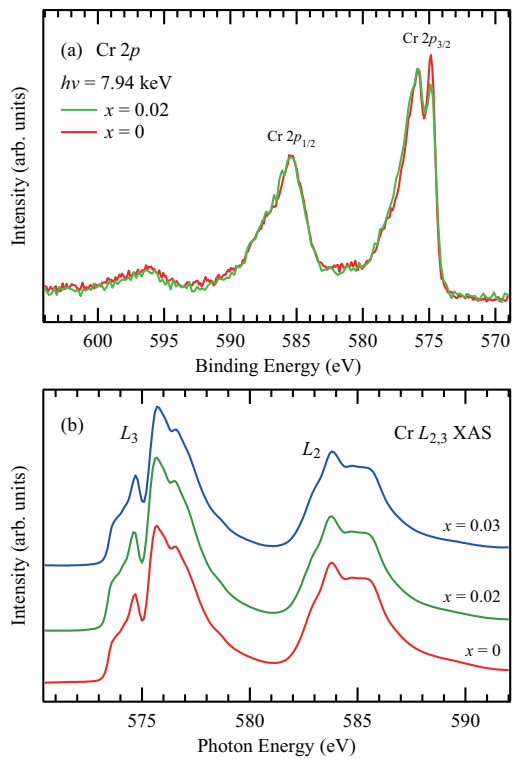}
	\caption{(Color) (a) Cr $2p$ core-level HX-PES spectra of CuCr$_{1-x}$Mg$x$O$_2$ ($x=$ 0, 0.02). Note that the background due to secondary electrons were subtracted by the Shirley method in order to analyze the intensity change with $x$. (b) Cr $L_{2,3}$ XAS spectra of CuCr$_{1-x}$Mg$x$O$_2$ ($x=$ 0, 0.02, 0.03).}
\label{FIG_Cr2p}
\end{center}
\end{figure}
%_________________________________________________________________________________________________________________________________

Figure \ref{FIG_Cr2p} shows Cr $2p$ core-level spectra of CuCrO$_2$ and CuCr$_{0.98}$Mg$_{0.02}$O$_2$.
The double-peak structure observed in the Cr $2p_{3/2}$ peak of the both samples is characteristic of Cr$^{3+}$
compound.
Both the spectra in Panel (a) are indeed very similar to those of Cr$_2$O$_3$ and CrN.\cite{Biesinger04,Bhobe10}
The $x=0$ and 0.02 spectra are very similar to each other, displaying Cr$^{3+}$ nature.
However, the  Cr $2p_{3/2}$ peak shows a remarkable change due to Mg doping; the first peak at 575 eV
obviously decreases in intensity with Mg doping.
A very similar change has recently been observed in CrN across its insulator-metal transition, which has been
explained by the screening effects due to mobile carriers.\cite{Bhobe10}
Therefore, the observed change is likely an evidence that doped holes move around the Cr sites, suggesting
the Cr $3d$ nature of a doped hole.
This result is consistent with the valence-band satellite analyses in Figs.\ \ref{FIG_VUV-Sat} and \ref{FIG_Cu2p3d}.
Nevertheless, all the three Cr $L_{2,3}$ XAS spectra in Fig.\ \ref{FIG_Cr2p}(b) are very similar to the reported
spectra of LaCrO$_3$ and Cr$_2$O$_3$,\cite{Sarma96,Matsubara02} indicating that the Cr ions are trivalent.
Unlike the Cu $L_3$ edge, Cr $L_{2,3}$ XAS spectra show no detectable changes with hole doping that were
observed for La$_{1-x}$Sr$_x$CrO$_3$ with $x\geq 0.3$.\cite{Sarma96}

\section{Discussion}

It is already established now that the ground-state electron configuration of Cu$_2$O is not a simple $|3d^{10}\rangle$,
but $\alpha|3d^{10}\rangle+\beta|3d^{9}4s\rangle$, while that of CuO is described as
$\alpha^\prime |3d^{9}\rangle+\beta^\prime |3d^{10}\underline{L}\rangle$, where $\underline{L}$ denotes an
O $2p$ ligand hole; the $d^{10}$ configuration of the Cu$^+$ ion should be spherical, but it was long ago
pointed out that the charge distribution in Cu$_2$O can be non-spherical due to the hybridization between
the $d_{3z^2-r^2}$ orbital ($z$ axis along the Cu-O bonding) and the $4s$ orbital,\cite{Orgel58} and has
been discussed theoretically later.\cite{Marksteiner86}
This hybridization yields a $d_{3z^2-r^2}$ hole and hence the ground state of  Cu$_2$O should have the
$|3d^{9}4s\rangle$ component.
The $d$ hole state has recently been directly observed,\cite{Zuo99} confirming the interpretation of the satellite
structures at 15 eV (the $|3d^{9}4s\rangle \to |3d^{8}4s\rangle$ process) in Cu$_2$O and at 13 eV
(the $|3d^{9}\rangle \to |3d^{8}\rangle$ process) in CuO.\cite{Thuler82,Ghijsen90,Shen90} 

The situation in CuCrO$_2$ is quite analogous to Cu$_2$O because the local environment around Cu is
the same O-Cu-O dumb-bell structure, and therefore it is not surprising that the ground state has the
$|3d^{9}4s\rangle$ component.
What is striking in our results is that even the $x=0$ sample with no Cu$^{2+}$ impurity centers has shown
a weak but detectable 13-eV satellite (Fig.~\ref{FIG_Cu2p3d}).
This inevitablly indicates that not a ``virtual'' $d^9$ state ($|3d^{9}4s\rangle$), but the ``real'' $d^9$ state
($|3d^{9}\rangle$) has to exist in CuCrO$_2$.
However, the Cu $2p$ core-level spectra do not show any trace of such a configuration even for hole-doped
samples, either.
Nevertheless, the development of the Cu $L_3$ pre-peak structure with $x$, again, undoubtedly demonstrates
that this $|3d^{9}\rangle$ configuration increases with $x$.
On the other hand, the doped hole should have the Cr $3d$ character from the Cr $2p$ HX-PES spectra while
the Cr $L$-edge XAS spectra show  no detectable changes.

To understand the above contradictory results, we reconsider the local electronic structure of the Cu site
beyond the nearest-neighbor oxygens, namely, consider the two metal sites, Cu and Cr, because their wave
functions are actually connected via the O $2p$ wave functions.

Within a metal-oxygen single cluster model (CuO$_2^{3-}$ and CrO$_6^{9-}$ for the Cu and the Cr sites,
respectively), the local electronic configuration of Cu$^{+}$ can be described as
$\alpha|3d^{10}\rangle+\beta|3d^{9}4s\rangle$, whereas that of Cr$^{3+}$ will be $\alpha^\prime|3d^{3}\rangle+\beta^\prime|3d^{4}\underline{L}\rangle$.\cite{Saitoh95,Uozumi97}
Although the $\underline{L}$ molecular orbitals of the Cu and the Cr sites have in fact different symmetries,
there should be sizable overlap between some of them as discussed in Fig.~\ref{FIG_VUV-VB}.
Hence, we consider the Cu-O-Cr cluster and re-define $\underline{L}$ as an O $2p$ ligand hole in a molecular
orbital of this cluster.
In this model, the combination of the $|3d^{9}4s\rangle$ configuration of Cu$^{+}$ and the
$|3d^{4}\underline{L}\rangle$ configuration of Cr$^{3+}$ can produce the $|3d^{9}\rangle$ and
$|3d^{4}\rangle$ configurations at the Cu and Cr sites, respectively, because of the extended nature
of the $4s$ state.
Hence, the ground state $|g\rangle$ can be described as 
\begin{equation*}
|g\rangle =
\alpha|d^{10} d^3 \rangle +
\beta|d^{9}\! s\, d^3 \rangle +
\gamma|d^{9} d^4 \rangle +
\delta|d^{10} \underline{L}\, d^4 \rangle ,
\end{equation*}
where the left $d^9$ and $d^{10}$ denote the Cu $3d$ states, $s$ denotes the Cu $4s$ state,
and the right $d^3$ and $d^4$ denote the Cr $3d$ states. $|d^{10} d^3 \rangle$ is the main
configuration, $|d^{9}\! s\, d^3 \rangle$ corresponds to the $d_{3z^2-r^2}$ hole state,
$|d^{9} d^4\rangle$ is the Cu $4s$--to--Cr $3d$ charge-transfer state, and finally
$|d^{10} \underline{L}\, d^4 \rangle$ originates from the O $2p$--to--Cr $3d$ charge-transfer state,
which is the second main configuration. The $|d^{9}\! s\, \underline{L}\, d^4 \rangle$ configuration
is not included because this is the origin of the $|d^{9} d^4 \rangle$ configuration.

The final state of the valence-band photoemission by Cu $3d$ emission is
\begin{equation*}
|f^{\rm{Cu}}_{\rm{v}} \rangle =
a|d^{10} \underline{L} \, d^3 \rangle +
b|d^{9} d^3 \rangle +
c|d^{8}\! s\, d^3 \rangle +
d|d^{8} d^4 \rangle .
\end{equation*}
Here, $|d^{9} \underline{L}\, d^4 \rangle$ is neglected because this configuration will easily
transform into $|d^{9} d^3 \rangle$ due to the combination of one extra electron at the Cr site
and the lack of one electron at the Cu site. 

For the Cu $2p$ core-level photoemission, the final state will be
\begin{equation*}
| f^{\rm{Cu}}_{\rm{c}} \rangle =
a|\underline{c}d^{10} d^3 \rangle +
b|\underline{c}d^{9}\! s\, d^3 \rangle +
c|\underline{c}d^{10} \underline{L}\, d^4 \rangle +
d|\underline{c}d^{9} d^4 \rangle ,
\end{equation*}
and for the Cu $L$-edge XAS, the final state will be
\begin{equation*}
| f^{\rm{Cu}}_L \rangle =
a|\underline{c}d^{10}\! s\, d^3 \rangle +
b|\underline{c}d^{10} d^4 \rangle +
c|\underline{c}d^{10}\! s\, \underline{L}\, d^4 \rangle ,
\end{equation*}
where $\underline{c}$ denotes a Cu $2p$ core hole.

Within this framework, the Cu $3p$-$3d$ and $2p$-$3d$ resonant photoemission spectra can have both
the $3d^84s$ (at 15 eV) and $3d^8$ (at 13 eV) final-state satellites due to the processes of
$|d^{9}\! s\, d^3 \rangle \to |d^{8}\! s\, d^3 \rangle$ and  $|d^9 d^4\rangle \to |d^8 d^4 \rangle$, respectively.
This scenario even predicts that CuAlO$_2$ will not have the 13-eV satellite because there are no available
Al states in the valence band, and indeed, an XPS spectrum of CuAlO$_2$ shows a dip around 13 eV, while
that of CuCrO$_2$ has extra spectral weight,\cite{Arnold09} supporting the scenario.
The absence of the $|\underline{c}3d^9\rangle$ final-state satellite in the Cu $2p$ core-level spectra can
be explained by strong screening effects due to the presence of a core hole at the Cu site:
The large $\underline{c}$-$d$ Coulomb attraction increases the number of $d$ electrons and accordingly
it makes the $|\underline{c}d^{9}\! s\, d^3 \rangle$ and $|\underline{c}d^{9} d^4 \rangle$ weight
negligible even for the lightly hole-doped samples.
Likewise, the lack of the pre-peak structure in the Cu $L$-edge XAS specturm of the $x$=0 sample can also
be explained by the core-hole screening effects that reduce the weight of the $|\underline{c}d^{10} d^4 \rangle$
(and the $|\underline{c}d^{10}\! s\, \underline{L}\, d^4 \rangle$) configuration(s) in $| f^{\rm{Cu}}_L \rangle$
(Fig.\ \ref{FIG_Cu2p3d}(b)).

%_________________________________________________________________________________________________________________________________
%Fig. 9
\begin{figure}[t]
    \begin{center}
    \includegraphics[width=67mm, keepaspectratio]{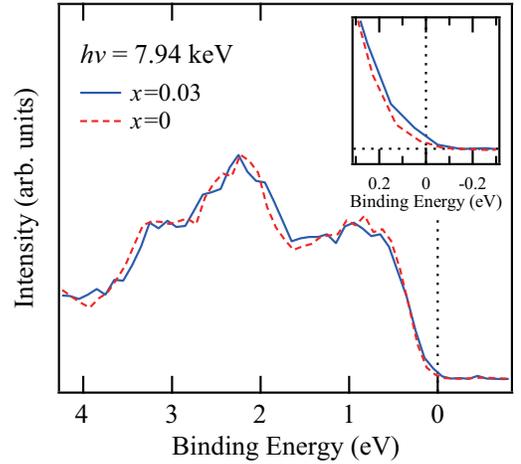}
    \caption{(Color online) Near-$E_F$ valence-band spectra of CuCr$_{1-x}$Mg$_x$O$_2$ ($x=$ 0, 0.03) taken with 7.94 keV. The intensity is normalized with respect to the spectral weight sum from $-0.5$ to 4.0 eV.}
    \label{FIG_NEFHXPES}
    \end{center}
    \end{figure}
%_________________________________________________________________________________________________________________________________

	From the above consideration, there must be weak but finite Cu $4s$ spectral weight at the top of the
valence band, and this can actually be observed; Figure \ref{FIG_NEFHXPES} shows a HX-PES valence-band
spectra of $x=0$ and 0.03 samples.
Considering that the photoionization cross section of $sp$ states of this energy range is largely enhanced,\cite{Yeh85}
the small enhancement at very near $E_F$ (see the inset) can be interpreted as an increase in the Cu $4sp$
emission due to hole doping, namely, supporting finite Cu $4s$ spectral weight at the top of the valence band.
It is accordingly revealed that the very top of the valence band has actually the Cu $4s$ character in addition to
the Cr $3d$ character.
This Cu $4s$--Cr $3d$ duality of a doped hole can explain the observed magnetic and transport properties of
CuCr$_{1-x}$Mg$_x$O$_2$; the doped holes moving in the Cr-O network can lift the magnetic frustration in
the Cr triangular spin lattice, resulting in an increase in the magnetic susceptibility with $x$.\cite{Okuda08}
The holes that are not restricted in the Cu-O network also explain a higher electric conductivity compared with
other hole-doped Cu delafossites such as CuAlO$_2$.\cite{Nagarajan01}
In particular, the highest conductivity by selecting Cr$^{3+}$ is strikingly demonstrating the importance
of the Cu-Cr combination.\cite{Nagarajan01}
From the viewpoint of the electronic structure, this can be interpreted as a consequence of an ``appropriate''
combination in terms of the difference of $\Delta_{\text{Cr}^{3+}}$ and $\Delta_{\text{Cr}^{+}}$ of
Cu$M$O$_2$.\cite{Iwasawa06}

%-------------------- CONCLUSIONS --------------------
\section{Conclusions}

We have studied the electronic structure of hole-doped delafossite oxides CuCr$_{1-x}$Mg$_x$O$_2$
by high-resolution photoemission spectroscopy, x-ray absorption spectroscopy, and LDA+$U$ band-structure
calculations.
The Cr and Cu $3p$-$3d$ resonant PES spectra demonstrated that the leading structure of the valence
band near the $E_F$ has primarily the Cr $3d$ character with a minor contribution from the Cu $3d$
due to hybridization with the O $2p$ states, in good agreement with the band-structure calculation.
This result indicates that a doped hole will primarily have the Cr $3d$ character.
The Cr $2p$ PES and $L$-edge XAS spectra of CuCr$_{1-x}$Mg$_x$O$_2$ showed typical Cr$^{3+}$ features,
whereas the Cu $L$-edge XAS spectra exhibited a systematic change with $x$.
This result, by contrast, indicates that the Cu valence is monovalent at $x$=0 and the holes will be doped into
the Cu sites, which contradicts the Cr and Cu $3p$-$3d$ resonant PES.
Nevertheless, the Cu $2p$-$3d$ resonant PES spectra display the two types of charge-transfer satellites that
should be attributed to Cu$^+$ ($3d^{10}$) and Cu$^{2+}$ ($3d^9$) like initial states, while the Cu $2p$
PES with no doubt shows the Cu$^{+}$ character even for $x>0$.

We have proposed that the above apparently contradictory results can consistently be understood by introducing
not only the Cu $4s$ state as traditionally, but also newly finite Cu $4s-$Cr $3d$ charge transfer via O $2p$ states
in the ground-state electronic configuration.
We found that this model can explain well some of the characteristic magnetic and transport properties of this compound.

\begin{acknowledgments}
The authors would like to thank T.\ Mizokawa for enlightening discussions.
The synchrotron radiation experiments at the Photon Factory and SPring-8 were performed under
the approval of the Photon Factory (Proposal Numbers 2008G688, 2010G655, 2009S2-005, and
2011S2-003) and of the Japan Synchrotron Radiation Research Institute (Proposal Numbers 2011A1624,
2011B1710, and 2012B1003), respectively.
This work was supported by JSPS KAKENHI Grants No.\ 22560786 and No.\ 23840039.
This work was also granted by JSPS the ``Funding Program for World-Leading Innovative R{\&}D
on Science and Technology (FIRST Program)'', initiated by the Council for Science and Technology
Policy (CSTP).
\end{acknowledgments}

%\bibliographystyle{apsrev4-1}
%\bibliography{CuCrO2refs}

\begin{thebibliography}{56}%
\makeatletter
\providecommand \@ifxundefined [1]{%
 \@ifx{#1\undefined}
}%
\providecommand \@ifnum [1]{%
 \ifnum #1\expandafter \@firstoftwo
 \else \expandafter \@secondoftwo
 \fi
}%
\providecommand \@ifx [1]{%
 \ifx #1\expandafter \@firstoftwo
 \else \expandafter \@secondoftwo
 \fi
}%
\providecommand \natexlab [1]{#1}%
\providecommand \enquote  [1]{``#1''}%
\providecommand \bibnamefont  [1]{#1}%
\providecommand \bibfnamefont [1]{#1}%
\providecommand \citenamefont [1]{#1}%
\providecommand \href@noop [0]{\@secondoftwo}%
\providecommand \href [0]{\begingroup \@sanitize@url \@href}%
\providecommand \@href[1]{\@@startlink{#1}\@@href}%
\providecommand \@@href[1]{\endgroup#1\@@endlink}%
\providecommand \@sanitize@url [0]{\catcode `\\12\catcode `\$12\catcode
  `\&12\catcode `\#12\catcode `\^12\catcode `\_12\catcode `\%12\relax}%
\providecommand \@@startlink[1]{}%
\providecommand \@@endlink[0]{}%
\providecommand \url  [0]{\begingroup\@sanitize@url \@url }%
\providecommand \@url [1]{\endgroup\@href {#1}{\urlprefix }}%
\providecommand \urlprefix  [0]{URL }%
\providecommand \Eprint [0]{\href }%
\providecommand \doibase [0]{http://dx.doi.org/}%
\providecommand \selectlanguage [0]{\@gobble}%
\providecommand \bibinfo  [0]{\@secondoftwo}%
\providecommand \bibfield  [0]{\@secondoftwo}%
\providecommand \translation [1]{[#1]}%
\providecommand \BibitemOpen [0]{}%
\providecommand \bibitemStop [0]{}%
\providecommand \bibitemNoStop [0]{.\EOS\space}%
\providecommand \EOS [0]{\spacefactor3000\relax}%
\providecommand \BibitemShut  [1]{\csname bibitem#1\endcsname}%
\let\auto@bib@innerbib\@empty
%</preamble>
\bibitem [{\citenamefont {Okuda}\ \emph {et~al.}(2005)\citenamefont {Okuda},
  \citenamefont {Jufuku}, \citenamefont {Hidaka},\ and\ \citenamefont
  {Terada}}]{Okuda05}%
  \BibitemOpen
  \bibfield  {author} {\bibinfo {author} {\bibfnamefont {T.}~\bibnamefont
  {Okuda}}, \bibinfo {author} {\bibfnamefont {N.}~\bibnamefont {Jufuku}},
  \bibinfo {author} {\bibfnamefont {S.}~\bibnamefont {Hidaka}},\ and\ \bibinfo
  {author} {\bibfnamefont {N.}~\bibnamefont {Terada}},\ }\href@noop {}
  {\bibfield  {journal} {\bibinfo  {journal} {Phys. Rev. B}\ }\textbf {\bibinfo
  {volume} {72}},\ \bibinfo {pages} {144403} (\bibinfo {year}
  {2005})}\BibitemShut {NoStop}%
\bibitem [{\citenamefont {Terasaki}\ \emph {et~al.}(1997)\citenamefont
  {Terasaki}, \citenamefont {Sasago},\ and\ \citenamefont
  {Uchinokura}}]{Terasaki97}%
  \BibitemOpen
  \bibfield  {author} {\bibinfo {author} {\bibfnamefont {I.}~\bibnamefont
  {Terasaki}}, \bibinfo {author} {\bibfnamefont {Y.}~\bibnamefont {Sasago}},\
  and\ \bibinfo {author} {\bibfnamefont {K.}~\bibnamefont {Uchinokura}},
  }\href@noop {} {\bibfield  {journal} {\bibinfo  {journal} {Phys. Rev. B}\
  }\textbf {\bibinfo {volume} {56}},\ \bibinfo {pages} {R12685} (\bibinfo
  {year} {1997})}\BibitemShut {NoStop}%
\bibitem [{\citenamefont {Takeuchi}\ \emph {et~al.}(2004)\citenamefont
  {Takeuchi}, \citenamefont {Kondo}, \citenamefont {Takami}, \citenamefont
  {Takahashi}, \citenamefont {Ikuta}, \citenamefont {Mizutani}, \citenamefont
  {Soda}, \citenamefont {Funahashi}, \citenamefont {Shikano}, \citenamefont
  {Mikami}, \citenamefont {Tsuda}, \citenamefont {Yokoya}, \citenamefont
  {Shin},\ and\ \citenamefont {Muro}}]{Takeuchi04}%
  \BibitemOpen
  \bibfield  {author} {\bibinfo {author} {\bibfnamefont {T.}~\bibnamefont
  {Takeuchi}}, \bibinfo {author} {\bibfnamefont {T.}~\bibnamefont {Kondo}},
  \bibinfo {author} {\bibfnamefont {T.}~\bibnamefont {Takami}}, \bibinfo
  {author} {\bibfnamefont {H.}~\bibnamefont {Takahashi}}, \bibinfo {author}
  {\bibfnamefont {H.}~\bibnamefont {Ikuta}}, \bibinfo {author} {\bibfnamefont
  {U.}~\bibnamefont {Mizutani}}, \bibinfo {author} {\bibfnamefont
  {K.}~\bibnamefont {Soda}}, \bibinfo {author} {\bibfnamefont {R.}~\bibnamefont
  {Funahashi}}, \bibinfo {author} {\bibfnamefont {M.}~\bibnamefont {Shikano}},
  \bibinfo {author} {\bibfnamefont {M.}~\bibnamefont {Mikami}}, \bibinfo
  {author} {\bibfnamefont {S.}~\bibnamefont {Tsuda}}, \bibinfo {author}
  {\bibfnamefont {T.}~\bibnamefont {Yokoya}}, \bibinfo {author} {\bibfnamefont
  {S.}~\bibnamefont {Shin}},\ and\ \bibinfo {author} {\bibfnamefont
  {T.}~\bibnamefont {Muro}},\ }\href@noop {} {\bibfield  {journal} {\bibinfo
  {journal} {Phys. Rev. B}\ }\textbf {\bibinfo {volume} {69}},\ \bibinfo
  {pages} {125410} (\bibinfo {year} {2004})}\BibitemShut {NoStop}%
\bibitem [{Note1()}]{Note1}%
  \BibitemOpen
  \bibinfo {note} {The doped hole amount is $x/(1-x)$}\BibitemShut {NoStop}%
\bibitem [{\citenamefont {Iwasawa}\ \emph {et~al.}(2006)\citenamefont
  {Iwasawa}, \citenamefont {Yamakawa}, \citenamefont {Saitoh}, \citenamefont
  {Inaba}, \citenamefont {Katsufuji}, \citenamefont {Higashiguchi},
  \citenamefont {Shimada}, \citenamefont {Namatame},\ and\ \citenamefont
  {Taniguchi}}]{Iwasawa06}%
  \BibitemOpen
  \bibfield  {author} {\bibinfo {author} {\bibfnamefont {H.}~\bibnamefont
  {Iwasawa}}, \bibinfo {author} {\bibfnamefont {K.}~\bibnamefont {Yamakawa}},
  \bibinfo {author} {\bibfnamefont {T.}~\bibnamefont {Saitoh}}, \bibinfo
  {author} {\bibfnamefont {J.}~\bibnamefont {Inaba}}, \bibinfo {author}
  {\bibfnamefont {T.}~\bibnamefont {Katsufuji}}, \bibinfo {author}
  {\bibfnamefont {M.}~\bibnamefont {Higashiguchi}}, \bibinfo {author}
  {\bibfnamefont {K.}~\bibnamefont {Shimada}}, \bibinfo {author} {\bibfnamefont
  {H.}~\bibnamefont {Namatame}},\ and\ \bibinfo {author} {\bibfnamefont
  {M.}~\bibnamefont {Taniguchi}},\ }\href@noop {} {\bibfield  {journal}
  {\bibinfo  {journal} {Phys. Rev. Lett.}\ }\textbf {\bibinfo {volume} {96}},\
  \bibinfo {pages} {067203} (\bibinfo {year} {2006})}\BibitemShut {NoStop}%
\bibitem [{\citenamefont {Kuroki}\ and\ \citenamefont
  {Arita}(2007)}]{Kuroki07}%
  \BibitemOpen
  \bibfield  {author} {\bibinfo {author} {\bibfnamefont {K.}~\bibnamefont
  {Kuroki}}\ and\ \bibinfo {author} {\bibfnamefont {R.}~\bibnamefont {Arita}},\
  }\href@noop {} {\bibfield  {journal} {\bibinfo  {journal} {J. Phys. Soc.
  Jpn.}\ }\textbf {\bibinfo {volume} {76}},\ \bibinfo {pages} {083707}
  (\bibinfo {year} {2007})}\BibitemShut {NoStop}%
\bibitem [{\citenamefont {Nagarajan}\ \emph {et~al.}(2001)\citenamefont
  {Nagarajan}, \citenamefont {Draeseke}, \citenamefont {Sleight},\ and\
  \citenamefont {Tate}}]{Nagarajan01}%
  \BibitemOpen
  \bibfield  {author} {\bibinfo {author} {\bibfnamefont {R.}~\bibnamefont
  {Nagarajan}}, \bibinfo {author} {\bibfnamefont {A.}~\bibnamefont {Draeseke}},
  \bibinfo {author} {\bibfnamefont {A.}~\bibnamefont {Sleight}}, \and\
  \bibinfo {author} {\bibfnamefont {J.}~\bibnamefont {Tate}},\ }\href@noop {}
  {\bibfield  {journal} {\bibinfo  {journal} {J. Appl. Phys.}\ }\textbf
  {\bibinfo {volume} {89}},\ \bibinfo {pages} {8022} (\bibinfo {year}
  {2001})}\BibitemShut {NoStop}%
\bibitem [{\citenamefont {Kimura}\ \emph {et~al.}(2006)\citenamefont {Kimura},
  \citenamefont {Lashley},\ and\ \citenamefont {Ramirez}}]{Kimura06}%
  \BibitemOpen
  \bibfield  {author} {\bibinfo {author} {\bibfnamefont {T.}~\bibnamefont
  {Kimura}}, \bibinfo {author} {\bibfnamefont {J.~C.}\ \bibnamefont {Lashley}},
  and\ \bibinfo {author} {\bibfnamefont {A.~P.}\ \bibnamefont {Ramirez}},\
  }\href@noop {} {\bibfield  {journal} {\bibinfo  {journal} {Phys. Rev. B}\
  }\textbf {\bibinfo {volume} {73}},\ \bibinfo {pages} {220401(R)} (\bibinfo
  {year} {2006})}\BibitemShut {NoStop}%
\bibitem [{\citenamefont {Seki}\ \emph {et~al.}(2008)\citenamefont {Seki},
  \citenamefont {Onose},\ and\ \citenamefont {Tokura}}]{Seki08}%
  \BibitemOpen
  \bibfield  {author} {\bibinfo {author} {\bibfnamefont {S.}~\bibnamefont
  {Seki}}, \bibinfo {author} {\bibfnamefont {Y.}~\bibnamefont {Onose}},\ and\
  \bibinfo {author} {\bibfnamefont {Y.}~\bibnamefont {Tokura}},\ }\href@noop {}
  {\bibfield  {journal} {\bibinfo  {journal} {Phys. Rev. Lett.}\ }\textbf
  {\bibinfo {volume} {101}},\ \bibinfo {pages} {067204} (\bibinfo {year}
  {2008})}\BibitemShut {NoStop}%
\bibitem [{\citenamefont {Hamberg}\ and\ \citenamefont
  {Granqvist}(1986)}]{Hamberg86}%
  \BibitemOpen
  \bibfield  {author} {\bibinfo {author} {\bibfnamefont {I.}~\bibnamefont
  {Hamberg}}\ and\ \bibinfo {author} {\bibfnamefont {C.~G.}\ \bibnamefont
  {Granqvist}},\ }\href@noop {} {\bibfield  {journal} {\bibinfo  {journal} {J.
  Appl. Phys.}\ }\textbf {\bibinfo {volume} {60}},\ \bibinfo {pages} {R123}
  (\bibinfo {year} {1986})}\BibitemShut {NoStop}%
\bibitem [{\citenamefont {Kawazoe}\ \emph {et~al.}(1997)\citenamefont
  {Kawazoe}, \citenamefont {Yasukawa}, \citenamefont {Hyodo}, \citenamefont
  {Kurita}, \citenamefont {Yanagi},\ and\ \citenamefont {Hosono}}]{Kawazoe97}%
  \BibitemOpen
  \bibfield  {author} {\bibinfo {author} {\bibfnamefont {H.}~\bibnamefont
  {Kawazoe}}, \bibinfo {author} {\bibfnamefont {M.}~\bibnamefont {Yasukawa}},
  \bibinfo {author} {\bibfnamefont {H.}~\bibnamefont {Hyodo}}, \bibinfo
  {author} {\bibfnamefont {M.}~\bibnamefont {Kurita}}, \bibinfo {author}
  {\bibfnamefont {H.}~\bibnamefont {Yanagi}},\ and\ \bibinfo {author}
  {\bibfnamefont {H.}~\bibnamefont {Hosono}},\ }\href@noop {} {\bibfield
  {journal} {\bibinfo  {journal} {Nature}\ }\textbf {\bibinfo {volume} {389}},\
  \bibinfo {pages} {939} (\bibinfo {year} {1997})}\BibitemShut {NoStop}%
\bibitem [{\citenamefont {Scanlon}\ \emph {et~al.}(2009)\citenamefont
  {Scanlon}, \citenamefont {Walsh}, \citenamefont {Morgan}, \citenamefont
  {Watson}, \citenamefont {Payne},\ and\ \citenamefont {Egdell}}]{Scanlon09}%
  \BibitemOpen
  \bibfield  {author} {\bibinfo {author} {\bibfnamefont {D.~O.}\ \bibnamefont
  {Scanlon}}, \bibinfo {author} {\bibfnamefont {A.}~\bibnamefont {Walsh}},
  \bibinfo {author} {\bibfnamefont {B.~J.}\ \bibnamefont {Morgan}}, \bibinfo
  {author} {\bibfnamefont {G.~W.}\ \bibnamefont {Watson}}, \bibinfo {author}
  {\bibfnamefont {D.~J.}\ \bibnamefont {Payne}},\ and\ \bibinfo {author}
  {\bibfnamefont {R.~G.}\ \bibnamefont {Egdell}},\ }\href@noop {} {\bibfield
  {journal} {\bibinfo  {journal} {Phys. Rev. B}\ }\textbf {\bibinfo {volume}
  {79}},\ \bibinfo {pages} {035101} (\bibinfo {year} {2009})}\BibitemShut
  {NoStop}%
\bibitem [{\citenamefont {Arnold}\ \emph {et~al.}(2009)\citenamefont {Arnold},
  \citenamefont {Payne}, \citenamefont {Bourlange}, \citenamefont {Hu},
  \citenamefont {Egdell}, \citenamefont {Piper}, \citenamefont {Colakerol},
  \citenamefont {De~Masi}, \citenamefont {Glans}, \citenamefont {Learmonth},
  \citenamefont {Smith}, \citenamefont {Guo}, \citenamefont {Scanlon},
  \citenamefont {Walsh}, \citenamefont {Morgan},\ and\ \citenamefont
  {Watson}}]{Arnold09}%
  \BibitemOpen
  \bibfield  {author} {\bibinfo {author} {\bibfnamefont {T.}~\bibnamefont
  {Arnold}}, \bibinfo {author} {\bibfnamefont {D.~J.}\ \bibnamefont {Payne}},
  \bibinfo {author} {\bibfnamefont {A.}~\bibnamefont {Bourlange}}, \bibinfo
  {author} {\bibfnamefont {J.~P.}\ \bibnamefont {Hu}}, \bibinfo {author}
  {\bibfnamefont {R.~G.}\ \bibnamefont {Egdell}}, \bibinfo {author}
  {\bibfnamefont {L.~F.~J.}\ \bibnamefont {Piper}}, \bibinfo {author}
  {\bibfnamefont {L.}~\bibnamefont {Colakerol}}, \bibinfo {author}
  {\bibfnamefont {A.}~\bibnamefont {De~Masi}}, \bibinfo {author} {\bibfnamefont
  {P.-A.}\ \bibnamefont {Glans}}, \bibinfo {author} {\bibfnamefont
  {T.}~\bibnamefont {Learmonth}}, \bibinfo {author} {\bibfnamefont {K.~E.}\
  \bibnamefont {Smith}}, \bibinfo {author} {\bibfnamefont {J.}~\bibnamefont
  {Guo}}, \bibinfo {author} {\bibfnamefont {D.~O.}\ \bibnamefont {Scanlon}},
  \bibinfo {author} {\bibfnamefont {A.}~\bibnamefont {Walsh}}, \bibinfo
  {author} {\bibfnamefont {B.~J.}\ \bibnamefont {Morgan}},\ and\ \bibinfo
  {author} {\bibfnamefont {G.~W.}\ \bibnamefont {Watson}},\ }\href@noop {}
  {\bibfield  {journal} {\bibinfo  {journal} {Phys. Rev. B}\ }\textbf {\bibinfo
  {volume} {79}},\ \bibinfo {pages} {075102} (\bibinfo {year}
  {2009})}\BibitemShut {NoStop}%
\bibitem [{\citenamefont {Hiraga}\ \emph {et~al.}(2011)\citenamefont {Hiraga},
  \citenamefont {Makino}, \citenamefont {Fukumura}, \citenamefont {Weng},\ and\
  \citenamefont {Kawasaki}}]{Hiraga11}%
  \BibitemOpen
  \bibfield  {author} {\bibinfo {author} {\bibfnamefont {H.}~\bibnamefont
  {Hiraga}}, \bibinfo {author} {\bibfnamefont {T.}~\bibnamefont {Makino}},
  \bibinfo {author} {\bibfnamefont {T.}~\bibnamefont {Fukumura}}, \bibinfo
  {author} {\bibfnamefont {H.}~\bibnamefont {Weng}},\ and\ \bibinfo {author}
  {\bibfnamefont {M.}~\bibnamefont {Kawasaki}},\ }\href@noop {} {\bibfield
  {journal} {\bibinfo  {journal} {Phys. Rev. B}\ }\textbf {\bibinfo {volume}
  {84}},\ \bibinfo {pages} {041411(R)} (\bibinfo {year} {2011})}\BibitemShut
  {NoStop}%
\bibitem [{\citenamefont {Maignan}\ \emph {et~al.}(2009)\citenamefont
  {Maignan}, \citenamefont {Martin}, \citenamefont {Fr\'{e}sard}, \citenamefont
  {Eyert}, \citenamefont {Guilmeau}, \citenamefont {H\'{e}bert}, \citenamefont
  {Poienar},\ and\ \citenamefont {Pelloquin}}]{Maignan09}%
  \BibitemOpen
  \bibfield  {author} {\bibinfo {author} {\bibfnamefont {A.}~\bibnamefont
  {Maignan}}, \bibinfo {author} {\bibfnamefont {C.}~\bibnamefont {Martin}},
  \bibinfo {author} {\bibfnamefont {R.}~\bibnamefont {Fr\'{e}sard}}, \bibinfo
  {author} {\bibfnamefont {V.}~\bibnamefont {Eyert}}, \bibinfo {author}
  {\bibfnamefont {E.}~\bibnamefont {Guilmeau}}, \bibinfo {author}
  {\bibfnamefont {S.}~\bibnamefont {H\'{e}bert}}, \bibinfo {author}
  {\bibfnamefont {M.}~\bibnamefont {Poienar}},\ and\ \bibinfo {author}
  {\bibfnamefont {D.}~\bibnamefont {Pelloquin}},\ }\href@noop {} {\bibfield
  {journal} {\bibinfo  {journal} {Solid State Commun.}\ }\textbf {\bibinfo
  {volume} {149}},\ \bibinfo {pages} {962} (\bibinfo {year}
  {2009})}\BibitemShut {NoStop}%
\bibitem [{\citenamefont {Ono}\ \emph {et~al.}(2007)\citenamefont {Ono},
  \citenamefont {Satoh}, \citenamefont {Nozaki},\ and\ \citenamefont
  {Kajitani}}]{Ono07}%
  \BibitemOpen
  \bibfield  {author} {\bibinfo {author} {\bibfnamefont {Y.}~\bibnamefont
  {Ono}}, \bibinfo {author} {\bibfnamefont {K.}~\bibnamefont {Satoh}}, \bibinfo
  {author} {\bibfnamefont {T.}~\bibnamefont {Nozaki}},\ and\ \bibinfo {author}
  {\bibfnamefont {T.}~\bibnamefont {Kajitani}},\ }\href@noop {} {\bibfield
  {journal} {\bibinfo  {journal} {Jpn. J. Appl. Phys.}\ }\textbf {\bibinfo
  {volume} {46}},\ \bibinfo {pages} {1071} (\bibinfo {year}
  {2007})}\BibitemShut {NoStop}%
\bibitem [{\citenamefont {Andersen}(1975)}]{Andersen75}%
  \BibitemOpen
  \bibfield  {author} {\bibinfo {author} {\bibfnamefont {O.~K.}\ \bibnamefont
  {Andersen}},\ }\href@noop {} {\bibfield  {journal} {\bibinfo  {journal}
  {Phys. Rev. B}\ }\textbf {\bibinfo {volume} {12}},\ \bibinfo {pages} {3060}
  (\bibinfo {year} {1975})}\BibitemShut {NoStop}%
\bibitem [{\citenamefont {Takeda}\ and\ \citenamefont
  {Kubler}(1979)}]{Takeda79}%
  \BibitemOpen
  \bibfield  {author} {\bibinfo {author} {\bibfnamefont {T.}~\bibnamefont
  {Takeda}}\ and\ \bibinfo {author} {\bibfnamefont {J.}~\bibnamefont
  {Kubler}},\ }\href@noop {} {\bibfield  {journal} {\bibinfo  {journal} {J.
  Phys. F: Met. Phys.}\ }\textbf {\bibinfo {volume} {9}},\ \bibinfo {pages}
  {661} (\bibinfo {year} {1979})}\BibitemShut {NoStop}%
\bibitem [{\citenamefont {Hohenberg}\ and\ \citenamefont
  {W.~Kohn}(1964)}]{Hohenberg64}%
  \BibitemOpen
  \bibfield  {author} {\bibinfo {author} {\bibfnamefont {P.}~\bibnamefont
  {Hohenberg}}\ and\ \bibinfo {author} {\bibfnamefont {W.}~\bibnamefont
  {W.~Kohn}},\ }\href@noop {} {\bibfield  {journal} {\bibinfo  {journal} {Phys.
  Rev.}\ }\textbf {\bibinfo {volume} {136}},\ \bibinfo {pages} {B864} (\bibinfo
  {year} {1964})}\BibitemShut {NoStop}%
\bibitem [{\citenamefont {Kohn}\ and\ \citenamefont {Sham}(1965)}]{Kohn65}%
  \BibitemOpen
  \bibfield  {author} {\bibinfo {author} {\bibfnamefont {W.}~\bibnamefont
  {Kohn}}\ and\ \bibinfo {author} {\bibfnamefont {L.~J.}\ \bibnamefont
  {Sham}},\ }\href@noop {} {\bibfield  {journal} {\bibinfo  {journal} {Phys.
  Rev.}\ }\textbf {\bibinfo {volume} {140}},\ \bibinfo {pages} {A1133}
  (\bibinfo {year} {1965})}\BibitemShut {NoStop}%
\bibitem [{\citenamefont {Vosko}\ \emph {et~al.}(1980)\citenamefont {Vosko},
  \citenamefont {Wilk},\ and\ \citenamefont {Nusair}}]{Vosko80}%
  \BibitemOpen
  \bibfield  {author} {\bibinfo {author} {\bibfnamefont {S.~H.}\ \bibnamefont
  {Vosko}}, \bibinfo {author} {\bibfnamefont {L.}~\bibnamefont {Wilk}},\ and\
  \bibinfo {author} {\bibfnamefont {M.}~\bibnamefont {Nusair}},\ }\href@noop {}
  {\bibfield  {journal} {\bibinfo  {journal} {Can. J. Phys.}\ }\textbf
  {\bibinfo {volume} {58}},\ \bibinfo {pages} {1200} (\bibinfo {year}
  {1980})}\BibitemShut {NoStop}%
\bibitem [{\citenamefont {Anisimov}\ \emph {et~al.}(1991)\citenamefont
  {Anisimov}, \citenamefont {Zaanen},\ and\ \citenamefont
  {Andersen}}]{Anisimov91}%
  \BibitemOpen
  \bibfield  {author} {\bibinfo {author} {\bibfnamefont {V.~I.}\ \bibnamefont
  {Anisimov}}, \bibinfo {author} {\bibfnamefont {J.}~\bibnamefont {Zaanen}},\
  and\ \bibinfo {author} {\bibfnamefont {O.~K.}\ \bibnamefont {Andersen}},\ 
  }\href@noop {} {\bibfield  {journal} {\bibinfo  {journal} {Phys. Rev. B}\
  }\textbf {\bibinfo {volume} {44}},\ \bibinfo {pages} {943} (\bibinfo {year}
  {1991})}\BibitemShut {NoStop}%
\bibitem [{\citenamefont {Solovyev}\ \emph {et~al.}(1996)\citenamefont
  {Solovyev}, \citenamefont {Hamada},\ and\ \citenamefont
  {Terakura}}]{Solovyev96}%
  \BibitemOpen
  \bibfield  {author} {\bibinfo {author} {\bibfnamefont {I.}~\bibnamefont
  {Solovyev}}, \bibinfo {author} {\bibfnamefont {N.}~\bibnamefont {Hamada}},\
  and\ \bibinfo {author} {\bibfnamefont {K.}~\bibnamefont {Terakura}},\
  }\href@noop {} {\bibfield  {journal} {\bibinfo  {journal} {Phys. Rev. B}\
  }\textbf {\bibinfo {volume} {53}},\ \bibinfo {pages} {7158} (\bibinfo {year}
  {1996})}\BibitemShut {NoStop}%
\bibitem [{\citenamefont {Anisimov}\ \emph {et~al.}(1997)\citenamefont
  {Anisimov}, \citenamefont {Aryasetiawan},\ and\ \citenamefont
  {Linchtenstein}}]{Anisimov97}%
  \BibitemOpen
  \bibfield  {author} {\bibinfo {author} {\bibfnamefont {V.~I.}\ \bibnamefont
  {Anisimov}}, \bibinfo {author} {\bibfnamefont {F.}~\bibnamefont
  {Aryasetiawan}},\ and\ \bibinfo {author} {\bibfnamefont {A.~I.}\
  \bibnamefont {Linchtenstein}},\ }\href@noop {} {\bibfield  {journal}
  {\bibinfo  {journal} {J. Phys.: Condens. Matter}\ }\textbf {\bibinfo {volume}
  {9}},\ \bibinfo {pages} {767} (\bibinfo {year} {1997})}\BibitemShut {NoStop}%
\bibitem [{\citenamefont {Poienar}\ \emph {et~al.}(2009)\citenamefont
  {Poienar}, \citenamefont {Damay}, \citenamefont {Martin}, \citenamefont
  {Hardy}, \citenamefont {Maignan},\ and\ \citenamefont {Andr\'e}}]{Poienar09}%
  \BibitemOpen
  \bibfield  {author} {\bibinfo {author} {\bibfnamefont {M.}~\bibnamefont
  {Poienar}}, \bibinfo {author} {\bibfnamefont {F.}~\bibnamefont {Damay}},
  \bibinfo {author} {\bibfnamefont {C.}~\bibnamefont {Martin}}, \bibinfo
  {author} {\bibfnamefont {V.}~\bibnamefont {Hardy}}, \bibinfo {author}
  {\bibfnamefont {A.}~\bibnamefont {Maignan}},\ and\ \bibinfo {author}
  {\bibfnamefont {G.}~\bibnamefont {Andr\'e}},\ }\href@noop {} {\bibfield
  {journal} {\bibinfo  {journal} {Phys. Rev. B}\ }\textbf {\bibinfo {volume}
  {79}},\ \bibinfo {pages} {014412} (\bibinfo {year} {2009})}\BibitemShut
  {NoStop}%
\bibitem [{\citenamefont {Li}\ \emph {et~al.}(1992)\citenamefont {Li},
  \citenamefont {Liu},\ and\ \citenamefont {Henrich}}]{Li92}%
  \BibitemOpen
  \bibfield  {author} {\bibinfo {author} {\bibfnamefont {X.}~\bibnamefont
  {Li}}, \bibinfo {author} {\bibfnamefont {L.}~\bibnamefont {Liu}},\ and\
  \bibinfo {author} {\bibfnamefont {V.~E.}\ \bibnamefont {Henrich}},\
  }\href@noop {} {\bibfield  {journal} {\bibinfo  {journal} {Solid State
  Commun.}\ }\textbf {\bibinfo {volume} {84}},\ \bibinfo {pages} {1103}
  (\bibinfo {year} {1992})}\BibitemShut {NoStop}%
\bibitem [{\citenamefont {Shirley}(1972)}]{Shirley72}%
  \BibitemOpen
  \bibfield  {author} {\bibinfo {author} {\bibfnamefont {D.~A.}\ \bibnamefont
  {Shirley}},\ }\href@noop {} {\bibfield  {journal} {\bibinfo  {journal} {Phys.
  Rev. B}\ }\textbf {\bibinfo {volume} {5}},\ \bibinfo {pages} {4709} (\bibinfo
  {year} {1972})}\BibitemShut {NoStop}%
\bibitem [{\citenamefont {Fano}(1961)}]{Fano61}%
  \BibitemOpen
  \bibfield  {author} {\bibinfo {author} {\bibfnamefont {U.}~\bibnamefont
  {Fano}},\ }\href@noop {} {\bibfield  {journal} {\bibinfo  {journal} {Phys.
  Rev.}\ }\textbf {\bibinfo {volume} {124}},\ \bibinfo {pages} {1866} (\bibinfo
  {year} {1961})}\BibitemShut {NoStop}%
\bibitem [{\citenamefont {Thuler}\ \emph {et~al.}(1982)\citenamefont {Thuler},
  \citenamefont {Benbow},\ and\ \citenamefont {Hurych}}]{Thuler82}%
  \BibitemOpen
  \bibfield  {author} {\bibinfo {author} {\bibfnamefont {M.~R.}\ \bibnamefont
  {Thuler}}, \bibinfo {author} {\bibfnamefont {R.~L.}\ \bibnamefont {Benbow}},
  and\ \bibinfo {author} {\bibfnamefont {Z.}~\bibnamefont {Hurych}},\
  }\href@noop {} {\bibfield  {journal} {\bibinfo  {journal} {Phys. Rev. B}\
  }\textbf {\bibinfo {volume} {26}},\ \bibinfo {pages} {669} (\bibinfo {year}
  {1982})}\BibitemShut {NoStop}%
\bibitem [{\citenamefont {Fujimori}\ \emph {et~al.}(1993)\citenamefont
  {Fujimori}, \citenamefont {Bocquet}, \citenamefont {Saitoh},\ and\
  \citenamefont {Mizokawa}}]{Fujimori93}%
  \BibitemOpen
  \bibfield  {author} {\bibinfo {author} {\bibfnamefont {A.}~\bibnamefont
  {Fujimori}}, \bibinfo {author} {\bibfnamefont {A.~E.}\ \bibnamefont
  {Bocquet}}, \bibinfo {author} {\bibfnamefont {T.}~\bibnamefont {Saitoh}},\
  and\ \bibinfo {author} {\bibfnamefont {T.}~\bibnamefont {Mizokawa}},\
  }\href@noop {} {\bibfield  {journal} {\bibinfo  {journal} {J. Electron
  Spectrosc. Relat. Phenomen.}\ }\textbf {\bibinfo {volume} {62}},\ \bibinfo
  {pages} {141} (\bibinfo {year} {1993})}\BibitemShut {NoStop}%
\bibitem [{\citenamefont {Saitoh}\ \emph {et~al.}(1995)\citenamefont {Saitoh},
  \citenamefont {Bocquet}, \citenamefont {Mizokawa},\ and\ \citenamefont
  {Fujimori}}]{Saitoh95}%
  \BibitemOpen
  \bibfield  {author} {\bibinfo {author} {\bibfnamefont {T.}~\bibnamefont
  {Saitoh}}, \bibinfo {author} {\bibfnamefont {A.~E.}\ \bibnamefont {Bocquet}},
  \bibinfo {author} {\bibfnamefont {T.}~\bibnamefont {Mizokawa}},\ and\
  \bibinfo {author} {\bibfnamefont {A.}~\bibnamefont {Fujimori}},\ }\href@noop
  {} {\bibfield  {journal} {\bibinfo  {journal} {Phys. Rev. B}\ }\textbf
  {\bibinfo {volume} {52}},\ \bibinfo {pages} {7934} (\bibinfo {year}
  {1995})}\BibitemShut {NoStop}%
\bibitem [{Note2()}]{Note2}%
  \BibitemOpen
  \bibinfo {note} {A $x=0.02$ spectrum is compared with theory because of a
  lower quality of $x=0.00$ spectra in this photon energy range.}\BibitemShut
  {Stop}%
\bibitem [{\citenamefont {Saitoh}\ \emph {et~al.}(1997)\citenamefont {Saitoh},
  \citenamefont {Mizokawa}, \citenamefont {Fujimori}, \citenamefont {Abbate},
  \citenamefont {Takeda},\ and\ \citenamefont {Takano}}]{Saitoh97}%
  \BibitemOpen
  \bibfield  {author} {\bibinfo {author} {\bibfnamefont {T.}~\bibnamefont
  {Saitoh}}, \bibinfo {author} {\bibfnamefont {T.}~\bibnamefont {Mizokawa}},
  \bibinfo {author} {\bibfnamefont {A.}~\bibnamefont {Fujimori}}, \bibinfo
  {author} {\bibfnamefont {M.}~\bibnamefont {Abbate}}, \bibinfo {author}
  {\bibfnamefont {Y.}~\bibnamefont {Takeda}},\ and\ \bibinfo {author}
  {\bibfnamefont {M.}~\bibnamefont {Takano}},\ }\href@noop {} {\bibfield
  {journal} {\bibinfo  {journal} {Phys. Rev. B}\ }\textbf {\bibinfo {volume}
  {55}},\ \bibinfo {pages} {4257} (\bibinfo {year} {1997})}\BibitemShut
  {NoStop}%
\bibitem [{\citenamefont {Iwasawa}\ \emph {et~al.}(2009)\citenamefont
  {Iwasawa}, \citenamefont {Kaneyoshi}, \citenamefont {Kurahashi},
  \citenamefont {Saitoh}, \citenamefont {Hase}, \citenamefont {Katsufuji},
  \citenamefont {Shimada}, \citenamefont {Namatame},\ and\ \citenamefont
  {Taniguchi}}]{Iwasawa09}%
  \BibitemOpen
  \bibfield  {author} {\bibinfo {author} {\bibfnamefont {H.}~\bibnamefont
  {Iwasawa}}, \bibinfo {author} {\bibfnamefont {S.}~\bibnamefont {Kaneyoshi}},
  \bibinfo {author} {\bibfnamefont {K.}~\bibnamefont {Kurahashi}}, \bibinfo
  {author} {\bibfnamefont {T.}~\bibnamefont {Saitoh}}, \bibinfo {author}
  {\bibfnamefont {I.}~\bibnamefont {Hase}}, \bibinfo {author} {\bibfnamefont
  {T.}~\bibnamefont {Katsufuji}}, \bibinfo {author} {\bibfnamefont
  {K.}~\bibnamefont {Shimada}}, \bibinfo {author} {\bibfnamefont
  {H.}~\bibnamefont {Namatame}},\ and\ \bibinfo {author} {\bibfnamefont
  {M.}~\bibnamefont {Taniguchi}},\ }\href@noop {} {\bibfield  {journal}
  {\bibinfo  {journal} {Phys. Rev. B}\ }\textbf {\bibinfo {volume} {80}},\
  \bibinfo {pages} {125122} (\bibinfo {year} {2009})}\BibitemShut {NoStop}%
\bibitem [{\citenamefont {Yeh}\ and\ \citenamefont {Lindau}(1985)}]{Yeh85}%
  \BibitemOpen
  \bibfield  {author} {\bibinfo {author} {\bibfnamefont {J.~J.}\ \bibnamefont
  {Yeh}}\ and\ \bibinfo {author} {\bibfnamefont {I.}~\bibnamefont {Lindau}},\
  }\href@noop {} {\bibfield  {journal} {\bibinfo  {journal} {At. Data Nucl.
  Data Tables}\ }\textbf {\bibinfo {volume} {32}},\ \bibinfo {pages} {1}
  (\bibinfo {year} {1985})}\BibitemShut {NoStop}%
\bibitem [{Note3()}]{Note3}%
  \BibitemOpen
  \bibinfo {note} {The binding energy ($E$) dependent Lorentzian FWHM was set
  to be $0.1(E-E_F)$ (eV).}\BibitemShut {Stop}%
\bibitem [{\citenamefont {Kizaki}\ \emph {et~al.}(2005)\citenamefont {Kizaki},
  \citenamefont {Sato}, \citenamefont {Yanase},\ and\ \citenamefont
  {Katayama-Yoshida}}]{Kizaki05}%
  \BibitemOpen
  \bibfield  {author} {\bibinfo {author} {\bibfnamefont {H.}~\bibnamefont
  {Kizaki}}, \bibinfo {author} {\bibfnamefont {K.}~\bibnamefont {Sato}},
  \bibinfo {author} {\bibfnamefont {A.}~\bibnamefont {Yanase}},\ and\ \bibinfo
  {author} {\bibfnamefont {H.}~\bibnamefont {Katayama-Yoshida}},\ }\href@noop
  {} {\bibfield  {journal} {\bibinfo  {journal} {Jpn. J. Appl. Phys.}\ }\textbf
  {\bibinfo {volume} {44}},\ \bibinfo {pages} {L1187} (\bibinfo {year}
  {2005})}\BibitemShut {NoStop}%
\bibitem [{\citenamefont {Ghijsen}\ \emph {et~al.}(1990)\citenamefont
  {Ghijsen}, \citenamefont {Tjeng}, \citenamefont {Eskes}, \citenamefont
  {Sawatzky},\ and\ \citenamefont {Johnson}}]{Ghijsen90}%
  \BibitemOpen
  \bibfield  {author} {\bibinfo {author} {\bibfnamefont {J.}~\bibnamefont
  {Ghijsen}}, \bibinfo {author} {\bibfnamefont {L.~H.}\ \bibnamefont {Tjeng}},
  \bibinfo {author} {\bibfnamefont {H.}~\bibnamefont {Eskes}}, \bibinfo
  {author} {\bibfnamefont {G.~A.}\ \bibnamefont {Sawatzky}},\ and\ \bibinfo
  {author} {\bibfnamefont {R.~L.}\ \bibnamefont {Johnson}},\ }\href@noop {}
  {\bibfield  {journal} {\bibinfo  {journal} {Phys. Rev. B}\ }\textbf {\bibinfo
  {volume} {42}},\ \bibinfo {pages} {2268} (\bibinfo {year}
  {1990})}\BibitemShut {NoStop}%
\bibitem [{\citenamefont {Shen}\ \emph {et~al.}(1990)\citenamefont {Shen},
  \citenamefont {List}, \citenamefont {Dessau}, \citenamefont {Parmigiani},
  \citenamefont {Arko}, \citenamefont {Bartlett}, \citenamefont {Wells},
  \citenamefont {Lindau},\ and\ \citenamefont {Spicer}}]{Shen90}%
  \BibitemOpen
  \bibfield  {author} {\bibinfo {author} {\bibfnamefont {Z.-X.}\ \bibnamefont
  {Shen}}, \bibinfo {author} {\bibfnamefont {R.~S.}\ \bibnamefont {List}},
  \bibinfo {author} {\bibfnamefont {D.~S.}\ \bibnamefont {Dessau}}, \bibinfo
  {author} {\bibfnamefont {F.}~\bibnamefont {Parmigiani}}, \bibinfo {author}
  {\bibfnamefont {A.~J.}\ \bibnamefont {Arko}}, \bibinfo {author}
  {\bibfnamefont {R.}~\bibnamefont {Bartlett}}, \bibinfo {author}
  {\bibfnamefont {B.~O.}\ \bibnamefont {Wells}}, \bibinfo {author}
  {\bibfnamefont {I.}~\bibnamefont {Lindau}},\ and\ \bibinfo {author}
  {\bibfnamefont {W.~E.}\ \bibnamefont {Spicer}},\ }\href@noop {} {\bibfield
  {journal} {\bibinfo  {journal} {Phys. Rev. B}\ }\textbf {\bibinfo {volume}
  {42}},\ \bibinfo {pages} {8081} (\bibinfo {year} {1990})}\BibitemShut
  {NoStop}%
\bibitem [{Note4()}]{Note4}%
  \BibitemOpen
  \bibinfo {note} {Because of a problem of energy calibration, our photon
  energy of the Cu $L_3$ XAS spectra is a little different from the reported
  values for CuCrO$_2$ or Cu$_2$O (Refs.~\protect \rev@citealpnum
  {Arnold09,Grioni89-1,Grioni92,Tjeng92}).}\BibitemShut {Stop}%
\bibitem [{\citenamefont {Tjeng}\ \emph {et~al.}(1992)\citenamefont {Tjeng},
  \citenamefont {Chen},\ and\ \citenamefont {Cheong}}]{Tjeng92}%
  \BibitemOpen
  \bibfield  {author} {\bibinfo {author} {\bibfnamefont {L.~H.}\ \bibnamefont
  {Tjeng}}, \bibinfo {author} {\bibfnamefont {C.~T.}\ \bibnamefont {Chen}},\
  and\ \bibinfo {author} {\bibfnamefont {S.-W.}\ \bibnamefont {Cheong}},\
  }\href@noop {} {\bibfield  {journal} {\bibinfo  {journal} {Phys. Rev. B}\
  }\textbf {\bibinfo {volume} {45}},\ \bibinfo {pages} {8205} (\bibinfo {year}
  {1992})}\BibitemShut {NoStop}%
\bibitem [{\citenamefont {Hulbert}\ \emph {et~al.}(1984)\citenamefont
  {Hulbert}, \citenamefont {Bunker}, \citenamefont {Brown},\ and\ \citenamefont
  {Pianetta}}]{Hulbert84}%
  \BibitemOpen
  \bibfield  {author} {\bibinfo {author} {\bibfnamefont {S.~L.}\ \bibnamefont
  {Hulbert}}, \bibinfo {author} {\bibfnamefont {B.~A.}\ \bibnamefont {Bunker}},
  \bibinfo {author} {\bibfnamefont {F.~C.}\ \bibnamefont {Brown}},\ and\
  \bibinfo {author} {\bibfnamefont {P.}~\bibnamefont {Pianetta}},\ }\href@noop
  {} {\bibfield  {journal} {\bibinfo  {journal} {Phys. Rev. B}\ }\textbf
  {\bibinfo {volume} {30}},\ \bibinfo {pages} {2120} (\bibinfo {year}
  {1984})}\BibitemShut {NoStop}%
\bibitem [{\citenamefont {Grioni}\ \emph {et~al.}(1989)\citenamefont {Grioni},
  \citenamefont {Goedkoop}, \citenamefont {Schoorl}, \citenamefont {de~Groot},
  \citenamefont {Fuggle}, \citenamefont {Sch\"afers}, \citenamefont {Koch},
  \citenamefont {Rossi}, \citenamefont {Esteva},\ and\ \citenamefont
  {Karnatak}}]{Grioni89-1}%
  \BibitemOpen
  \bibfield  {author} {\bibinfo {author} {\bibfnamefont {M.}~\bibnamefont
  {Grioni}}, \bibinfo {author} {\bibfnamefont {J.~B.}\ \bibnamefont
  {Goedkoop}}, \bibinfo {author} {\bibfnamefont {R.}~\bibnamefont {Schoorl}},
  \bibinfo {author} {\bibfnamefont {F.~M.~F.}\ \bibnamefont {de~Groot}},
  \bibinfo {author} {\bibfnamefont {J.~C.}\ \bibnamefont {Fuggle}}, \bibinfo
  {author} {\bibfnamefont {F.}~\bibnamefont {Sch\"afers}}, \bibinfo {author}
  {\bibfnamefont {E.~E.}\ \bibnamefont {Koch}}, \bibinfo {author}
  {\bibfnamefont {G.}~\bibnamefont {Rossi}}, \bibinfo {author} {\bibfnamefont
  {J.-M.}\ \bibnamefont {Esteva}},\ and\ \bibinfo {author} {\bibfnamefont
  {R.~C.}\ \bibnamefont {Karnatak}},\ }\href@noop {} {\bibfield  {journal}
  {\bibinfo  {journal} {Phys. Rev. B}\ }\textbf {\bibinfo {volume} {39}},\
  \bibinfo {pages} {1541} (\bibinfo {year} {1989})}\BibitemShut {NoStop}%
\bibitem [{\citenamefont {Aston}\ \emph {et~al.}(2005)\citenamefont {Aston},
  \citenamefont {Payne}, \citenamefont {Green}, \citenamefont {Egdell},
  \citenamefont {Law}, \citenamefont {Guo}, \citenamefont {Glans},
  \citenamefont {Learmonth},\ and\ \citenamefont {Smith}}]{Aston05}%
  \BibitemOpen
  \bibfield  {author} {\bibinfo {author} {\bibfnamefont {D.~J.}\ \bibnamefont
  {Aston}}, \bibinfo {author} {\bibfnamefont {D.~J.}\ \bibnamefont {Payne}},
  \bibinfo {author} {\bibfnamefont {A.~J.~H.}\ \bibnamefont {Green}}, \bibinfo
  {author} {\bibfnamefont {R.~G.}\ \bibnamefont {Egdell}}, \bibinfo {author}
  {\bibfnamefont {D.~S.~L.}\ \bibnamefont {Law}}, \bibinfo {author}
  {\bibfnamefont {J.}~\bibnamefont {Guo}}, \bibinfo {author} {\bibfnamefont
  {P.~A.}\ \bibnamefont {Glans}}, \bibinfo {author} {\bibfnamefont
  {T.}~\bibnamefont {Learmonth}},\ and\ \bibinfo {author} {\bibfnamefont
  {K.~E.}\ \bibnamefont {Smith}},\ }\href@noop {} {\bibfield  {journal}
  {\bibinfo  {journal} {Phys. Rev. B}\ }\textbf {\bibinfo {volume} {72}},\
  \bibinfo {pages} {195115} (\bibinfo {year} {2005})}\BibitemShut {NoStop}%
\bibitem [{\citenamefont {Grioni}\ \emph {et~al.}(1992)\citenamefont {Grioni},
  \citenamefont {van Acker}, \citenamefont {Czyzyk},\ and\ \citenamefont
  {Fuggle}}]{Grioni92}%
  \BibitemOpen
  \bibfield  {author} {\bibinfo {author} {\bibfnamefont {M.}~\bibnamefont
  {Grioni}}, \bibinfo {author} {\bibfnamefont {J.~F.}\ \bibnamefont {van
  Acker}}, \bibinfo {author} {\bibfnamefont {M.~T.}\ \bibnamefont {Czyzyk}},\
  and\ \bibinfo {author} {\bibfnamefont {J.~C.}\ \bibnamefont {Fuggle}},\
  }\href@noop {} {\bibfield  {journal} {\bibinfo  {journal} {Phys. Rev. B}\
  }\textbf {\bibinfo {volume} {45}},\ \bibinfo {pages} {3309} (\bibinfo {year}
  {1992})}\BibitemShut {NoStop}%
\bibitem [{\citenamefont {Le}\ \emph {et~al.}(2011)\citenamefont {Le},
  \citenamefont {Flahaut}, \citenamefont {Martinez}, \citenamefont {Andreu},
  \citenamefont {Gonbeau}, \citenamefont {Pachoud}, \citenamefont {Pelloquin},\
  and\ \citenamefont {Maignan}}]{Le11}%
  \BibitemOpen
  \bibfield  {author} {\bibinfo {author} {\bibfnamefont {T.~K.}\ \bibnamefont
  {Le}}, \bibinfo {author} {\bibfnamefont {D.}~\bibnamefont {Flahaut}},
  \bibinfo {author} {\bibfnamefont {H.}~\bibnamefont {Martinez}}, \bibinfo
  {author} {\bibfnamefont {N.}~\bibnamefont {Andreu}}, \bibinfo {author}
  {\bibfnamefont {D.}~\bibnamefont {Gonbeau}}, \bibinfo {author} {\bibfnamefont
  {E.}~\bibnamefont {Pachoud}}, \bibinfo {author} {\bibfnamefont
  {D.}~\bibnamefont {Pelloquin}},\ and\ \bibinfo {author} {\bibfnamefont
  {A.}~\bibnamefont {Maignan}},\ }\href@noop {} {\bibfield  {journal} {\bibinfo
   {journal} {J. Solid State Chem.}\ }\textbf {\bibinfo {volume} {184}},\
  \bibinfo {pages} {2387} (\bibinfo {year} {2011})}\BibitemShut {NoStop}%
\bibitem [{\citenamefont {Doniach}\ and\ \citenamefont
  {\v{S}unji\'c}(1970)}]{Doniach70}%
  \BibitemOpen
  \bibfield  {author} {\bibinfo {author} {\bibfnamefont {S.}~\bibnamefont
  {Doniach}}\ and\ \bibinfo {author} {\bibfnamefont {M.}~\bibnamefont
  {\v{S}unji\'c}},\ }\href@noop {} {\bibfield  {journal} {\bibinfo  {journal}
  {J. Phys. C}\ }\textbf {\bibinfo {volume} {3}},\ \bibinfo {pages} {285}
  (\bibinfo {year} {1970})}\BibitemShut {NoStop}%
\bibitem [{Note5()}]{Note5}%
  \BibitemOpen
  \bibinfo {note} {The derived asymmetric parameter $\alpha $ was $1.5\times
  10^{-5}$ ($x=0$) and $1.0\times 10^{-2}$ ($x=0.03$).}\BibitemShut {Stop}%
\bibitem [{\citenamefont {Biesinger}\ \emph {et~al.}(2004)\citenamefont
  {Biesinger}, \citenamefont {Brown}, \citenamefont {Mycroft}, \citenamefont
  {Davidson},\ and\ \citenamefont {McIntyre}}]{Biesinger04}%
  \BibitemOpen
  \bibfield  {author} {\bibinfo {author} {\bibfnamefont {M.~C.}\ \bibnamefont
  {Biesinger}}, \bibinfo {author} {\bibfnamefont {C.}~\bibnamefont {Brown}},
  \bibinfo {author} {\bibfnamefont {J.~R.}\ \bibnamefont {Mycroft}}, \bibinfo
  {author} {\bibfnamefont {R.~D.}\ \bibnamefont {Davidson}},\ and\ \bibinfo
  {author} {\bibfnamefont {N.~S.}\ \bibnamefont {McIntyre}},\ }\href@noop {}
  {\bibfield  {journal} {\bibinfo  {journal} {Surf. Interface Anal.}\ }\textbf
  {\bibinfo {volume} {36}},\ \bibinfo {pages} {1550} (\bibinfo {year}
  {2004})}\BibitemShut {NoStop}%
\bibitem [{\citenamefont {Bhobe}\ \emph {et~al.}(2010)\citenamefont {Bhobe},
  \citenamefont {Chainani}, \citenamefont {Taguchi}, \citenamefont {Takeuchi},
  \citenamefont {Eguchi}, \citenamefont {Matsunami}, \citenamefont {Ishizaka},
  \citenamefont {Takata}, \citenamefont {Oura}, \citenamefont {Senba},
  \citenamefont {Ohashi}, \citenamefont {Nishino}, \citenamefont {Yabashi},
  \citenamefont {Tamasaku}, \citenamefont {Ishikawa}, \citenamefont {Takenaka},
  \citenamefont {Takagi},\ and\ \citenamefont {Shin}}]{Bhobe10}%
  \BibitemOpen
  \bibfield  {author} {\bibinfo {author} {\bibfnamefont {P.~A.}\ \bibnamefont
  {Bhobe}}, \bibinfo {author} {\bibfnamefont {A.}~\bibnamefont {Chainani}},
  \bibinfo {author} {\bibfnamefont {M.}~\bibnamefont {Taguchi}}, \bibinfo
  {author} {\bibfnamefont {T.}~\bibnamefont {Takeuchi}}, \bibinfo {author}
  {\bibfnamefont {R.}~\bibnamefont {Eguchi}}, \bibinfo {author} {\bibfnamefont
  {M.}~\bibnamefont {Matsunami}}, \bibinfo {author} {\bibfnamefont
  {K.}~\bibnamefont {Ishizaka}}, \bibinfo {author} {\bibfnamefont
  {Y.}~\bibnamefont {Takata}}, \bibinfo {author} {\bibfnamefont
  {M.}~\bibnamefont {Oura}}, \bibinfo {author} {\bibfnamefont {Y.}~\bibnamefont
  {Senba}}, \bibinfo {author} {\bibfnamefont {H.}~\bibnamefont {Ohashi}},
  \bibinfo {author} {\bibfnamefont {Y.}~\bibnamefont {Nishino}}, \bibinfo
  {author} {\bibfnamefont {M.}~\bibnamefont {Yabashi}}, \bibinfo {author}
  {\bibfnamefont {K.}~\bibnamefont {Tamasaku}}, \bibinfo {author}
  {\bibfnamefont {T.}~\bibnamefont {Ishikawa}}, \bibinfo {author}
  {\bibfnamefont {K.}~\bibnamefont {Takenaka}}, \bibinfo {author}
  {\bibfnamefont {H.}~\bibnamefont {Takagi}},\ and\ \bibinfo {author}
  {\bibfnamefont {S.}~\bibnamefont {Shin}},\ }\href@noop {} {\bibfield
  {journal} {\bibinfo  {journal} {Phys. Rev. Lett.}\ }\textbf {\bibinfo
  {volume} {104}},\ \bibinfo {pages} {236404} (\bibinfo {year}
  {2010})}\BibitemShut {NoStop}%
\bibitem [{\citenamefont {Sarma}\ \emph {et~al.}(1996)\citenamefont {Sarma},
  \citenamefont {Maiti}, \citenamefont {Vescovo}, \citenamefont {Carbone},
  \citenamefont {Eberhardt}, \citenamefont {Rader},\ and\ \citenamefont
  {Gudat}}]{Sarma96}%
  \BibitemOpen
  \bibfield  {author} {\bibinfo {author} {\bibfnamefont {D.~D.}\ \bibnamefont
  {Sarma}}, \bibinfo {author} {\bibfnamefont {K.}~\bibnamefont {Maiti}},
  \bibinfo {author} {\bibfnamefont {E.}~\bibnamefont {Vescovo}}, \bibinfo
  {author} {\bibfnamefont {C.}~\bibnamefont {Carbone}}, \bibinfo {author}
  {\bibfnamefont {W.}~\bibnamefont {Eberhardt}}, \bibinfo {author}
  {\bibfnamefont {O.}~\bibnamefont {Rader}},\ and\ \bibinfo {author}
  {\bibfnamefont {W.}~\bibnamefont {Gudat}},\ }\href@noop {} {\bibfield
  {journal} {\bibinfo  {journal} {Phys. Rev. B}\ }\textbf {\bibinfo {volume}
  {53}},\ \bibinfo {pages} {13369} (\bibinfo {year} {1996})}\BibitemShut
  {NoStop}%
\bibitem [{\citenamefont {Matsubara}\ \emph {et~al.}(2002)\citenamefont
  {Matsubara}, \citenamefont {Uozumi}, \citenamefont {Kotani}, \citenamefont
  {Harada},\ and\ \citenamefont {Shin}}]{Matsubara02}%
  \BibitemOpen
  \bibfield  {author} {\bibinfo {author} {\bibfnamefont {M.}~\bibnamefont
  {Matsubara}}, \bibinfo {author} {\bibfnamefont {T.}~\bibnamefont {Uozumi}},
  \bibinfo {author} {\bibfnamefont {A.}~\bibnamefont {Kotani}}, \bibinfo
  {author} {\bibfnamefont {Y.}~\bibnamefont {Harada}},\ and\ \bibinfo {author}
  {\bibfnamefont {S.}~\bibnamefont {Shin}},\ }\href@noop {} {\bibfield
  {journal} {\bibinfo  {journal} {J. Phys. Soc. Jpn.}\ }\textbf {\bibinfo
  {volume} {71}},\ \bibinfo {pages} {347} (\bibinfo {year} {2002})}\BibitemShut
  {NoStop}%
\bibitem [{\citenamefont {Orgel}(1958)}]{Orgel58}%
  \BibitemOpen
  \bibfield  {author} {\bibinfo {author} {\bibfnamefont {L.~E.}\ \bibnamefont
  {Orgel}},\ }\href@noop {} {\bibfield  {journal} {\bibinfo  {journal} {J.
  Chem. Soc.}\ \bibinfo {pages} {4186}} (\bibinfo {year}
  {1958})}\BibitemShut {NoStop}%
\bibitem [{\citenamefont {Marksteiner}\ \emph {et~al.}(1986)\citenamefont
  {Marksteiner}, \citenamefont {Blaha},\ and\ \citenamefont
  {Schwarz}}]{Marksteiner86}%
  \BibitemOpen
  \bibfield  {author} {\bibinfo {author} {\bibfnamefont {P.}~\bibnamefont
  {Marksteiner}}, \bibinfo {author} {\bibfnamefont {P.}~\bibnamefont {Blaha}},
  and\ \bibinfo {author} {\bibfnamefont {K.}~\bibnamefont {Schwarz}},\
  }\href@noop {} {\bibfield  {journal} {\bibinfo  {journal} {Z. Phys. B}\
  }\textbf {\bibinfo {volume} {64}},\ \bibinfo {pages} {119} (\bibinfo {year}
  {1986})}\BibitemShut {NoStop}%
\bibitem [{\citenamefont {Zuo}\ \emph {et~al.}(1999)\citenamefont {Zuo},
  \citenamefont {Kim}, \citenamefont {O'Keeffe},\ and\ \citenamefont
  {Spence}}]{Zuo99}%
  \BibitemOpen
  \bibfield  {author} {\bibinfo {author} {\bibfnamefont {J.~M.}\ \bibnamefont
  {Zuo}}, \bibinfo {author} {\bibfnamefont {M.}~\bibnamefont {Kim}}, \bibinfo
  {author} {\bibfnamefont {M.}~\bibnamefont {O'Keeffe}},\ and\ \bibinfo
  {author} {\bibfnamefont {J.~C.~H.}\ \bibnamefont {Spence}},\ }\href@noop {}
  {\bibfield  {journal} {\bibinfo  {journal} {Nature}\ }\textbf {\bibinfo
  {volume} {401}},\ \bibinfo {pages} {49} (\bibinfo {year} {1999})}\BibitemShut
  {NoStop}%
\bibitem [{\citenamefont {Uozumi}\ \emph {et~al.}(1997)\citenamefont {Uozumi},
  \citenamefont {Okada}, \citenamefont {Kotani}, \citenamefont {Zimmermann},
  \citenamefont {Steiner}, \citenamefont {H\"ufner}, \citenamefont {Tezuka},\
  and\ \citenamefont {Shin}}]{Uozumi97}%
  \BibitemOpen
  \bibfield  {author} {\bibinfo {author} {\bibfnamefont {T.}~\bibnamefont
  {Uozumi}}, \bibinfo {author} {\bibfnamefont {K.}~\bibnamefont {Okada}},
  \bibinfo {author} {\bibfnamefont {A.}~\bibnamefont {Kotani}}, \bibinfo
  {author} {\bibfnamefont {R.}~\bibnamefont {Zimmermann}}, \bibinfo {author}
  {\bibfnamefont {P.}~\bibnamefont {Steiner}}, \bibinfo {author} {\bibfnamefont
  {S.}~\bibnamefont {H\"ufner}}, \bibinfo {author} {\bibfnamefont
  {Y.}~\bibnamefont {Tezuka}},\ and\ \bibinfo {author} {\bibfnamefont
  {S.}~\bibnamefont {Shin}},\ }\href@noop {} {\bibfield  {journal} {\bibinfo
  {journal} {J. Electron Spectrsc. Relat. Phenom.}\ }\textbf {\bibinfo
  {volume} {83}},\ \bibinfo {pages} {9} (\bibinfo {year} {1997})}\BibitemShut
  {NoStop}%
\bibitem{Okuda08}
	T. Okuda, Y. Beppu, Y. Fujii, T. Onoe, N .Terada, and S. Miyasaka, Phys. Rev. B
	\textbf{77}, 134423 (2008); T. Okuda, R. Kajimoto, M. Okawa, and T. Saitoh,
	Int. J. Mod. Phys. B \textbf{27}, 1330002 (2013).
\end{thebibliography}


\end{document}